\documentclass[aps,superscriptaddress,prc,twocolumn,nofootinbib]{revtex4}

\usepackage{amsmath,bm}
\usepackage{graphicx}
\usepackage{epstopdf}
\usepackage{subfigure}
\usepackage{epsfig}
\usepackage{amsmath,amssymb,amsfonts}
\usepackage{color}
\usepackage[utf8]{inputenc}
\usepackage{verbatim}
\usepackage{array}
\usepackage{hyperref}
\begin{document}

\renewcommand{\topfraction}{0.85}
\renewcommand{\textfraction}{0.1}
\renewcommand{\floatpagefraction}{0.75}

\title{Parton splitting scales of reclustered large-radius jets  in high-energy nuclear collisions}

\date{\today  \hspace{1ex}}

\author{Shan-Liang Zhang}
\affiliation{Guangdong Provincial Key Laboratory of Nuclear Science, Institute of Quantum Matter, South China Normal University, Guangzhou 510006, China.}
\affiliation{Key Laboratory of Quark \& Lepton Physics (MOE) and Institute of Particle Physics,
 Central China Normal University, Wuhan 430079, China}
\affiliation{Guangdong-Hong Kong Joint Laboratory of Quantum Matter, Southern Nuclear Science Computing Center, South China Normal University, Guangzhou 510006, China.}

\author{Meng-Quan Yang}
\affiliation{Key Laboratory of Quark \& Lepton Physics (MOE) and Institute of Particle Physics,
 Central China Normal University, Wuhan 430079, China}

\author{Ben-Wei Zhang \footnote{bwzhang@mail.ccnu.edu.cn}}
\affiliation{Key Laboratory of Quark \& Lepton Physics (MOE) and Institute of Particle Physics,
 Central China Normal University, Wuhan 430079, China}
\affiliation{Guangdong Provincial Key Laboratory of Nuclear Science, Institute of Quantum Matter, South China Normal University, Guangzhou 510006, China.}

\begin{abstract}
    We carry out the first theoretical investigation on yields and the hardest parton splitting of large-radius jets reclustered from small radius ($R=0.2$) anti-$k_t$ jets in Pb+Pb collisions, and confront them with the recent ATLAS measurements.
  The Linear Boltzmann Transport (LBT) model is employed for  jet propagation and jet-induced medium excitation in the hot-dense medium. We demonstrate that, with their complex structures, the medium suppression of the reclustered large radius jets at $R=1$ is larger than that of inclusive $R=0.4$ jets defined conventionally. The large radius jet constituents are reclustered  with the $k_t$ algorithm to obtain the splitting scale $\sqrt{d_{12}}$, which characterizes the transverse momentum scale for the hardest splitting in the jet. The large-radius jet production as a function of the splitting scale $\sqrt{d_{12}}$ of the hardest parton splitting is overall suppressed in Pb+Pb  relative to p+p collisions due to the reduction of jets yields. A detailed analyses show that the alterations of jet substructures in Pb+Pb also make significant contribution to the splitting scale $\sqrt{d_{12}}$ dependence of the nuclear modification factor $R_{AA}$. Numerical results for the medium modifications of the jet splitting angle $\Delta R_{12}$ and the splitting fraction $z$ are also presented.

\end{abstract}


\maketitle

\section{introduction}
\label{sec:Intro}


Jet quenching has long been proposed as an excellent probe of the properties of the Quark-Gluon Plasma(QGP) \cite{Wang:1991xy,Wang:1998ww}, formed in high-energy heavy-ion collisions,  such as those experiments at  the Large Hadron Collider (LHC) \cite{Aad:2010bu,Chatrchyan:2012gt,Chatrchyan:2013kwa,Aad:2014bxa} and the Relativistic Heavy Ion Collider(RHIC)\cite{Adcox:2001jp,Adler:2002xw, Adler:2002tq}. Parton energy loss in the dense QCD medium can lead to suppression of the single inclusive hadron spectra at large transverse momentum \cite{Adcox:2001jp,Adler:2002xw, Adler:2002tq,Qin:2007rn,Chen:2011vt,Buzzatti:2011vt,Majumder:2011uk,Aamodt:2010jd,CMS:2012aa},
 shift of $\gamma$-hadron \cite{Wang:1996yh,Renk:2006qg,Zhang:2009rn,Qin:2009bk,Adare:2009vd,Abelev:2009gu,Chen:2017zte} and dihadron
 transverse momentum asymmetry~\cite{Zhang:2007ja,stardihadron,Ayala:2009fe,Ayala:2011ii} in high-energy heavy-ion collisions.  Phenomenological studies of experimental data on these observables at RHIC and LHC have provided important constraints on the properties of QGP.  Jet transport coefficient $\hat{q}$ indicating the transverse momentum
broadening squared per unit distance due to jet-medium interaction was extracted from a set of single
inclusive hadron spectra by JET Collaboration \cite{Burke:2013yra} and from other observables \cite{Dai:2017piq,Dai:2017tuy,Ma:2018swx, Xie:2019oxg,Andres:2016iys,JETSCAPE:2021ehl,JETSCAPE:2020shq} .

The fragmentations of large transverse momentum
($p_T$) quarks or gluons result in groups of angularly-correlated particles, referred to as jets, which pave a new way to study the interactions between jet partons and the medium constituents. Production and suppression of fully reconstructed single jets~\cite{Aad:2014bxa,Adam:2015ewa,Khachatryan:2016jfl,Wang:2016fds,Chien:2015hda,Casalderrey-Solana:2014bpa,Tachibana:2017syd,Chen:2019gqo,Yan:2020zrz,He:2020iow,He:2018gks,CMS:2012ulu,ALICE:2013dpt}, di-jets~\cite{Aad:2010bu,Chatrchyan:2011sx,Vitev:2009rd,Qin:2010mn,Young:2011qx,He:2011pd,Renk:2012cx} and $\gamma/Z^0$-jets~\cite{Dai:2012am,Neufeld:2010fj,Kang:2017xnc,Chatrchyan:2012gt,Luo:2018pto,Zhang:2018urd,Sirunyan:2017jic,Yang:2021qtl} have also been studied in heavy-ion
collisions and they can provide additional constraints on the jet-medium interaction and jet transport coefficient.
Investigations on the jet substructures: jet shape~\cite{Vitev:2008rz,Chatrchyan:2013kwa,Sirunyan:2018ncy,Chang:2019sae,KunnawalkamElayavalli:2017hxo,Kang:2017mda,Ma:2013uqa}, jet fragmentation functions~\cite{Chatrchyan:2014ava,Aaboud:2017bzv,Sirunyan:2018qec,Aaboud:2019oac,Chen:2020tbl}, groomed subjets~\cite{Sirunyan:2017bsd,Sirunyan:2018gct,Acharya:2019djg,Andrews:2018jcm,Casalderrey-Solana:2019ubu}, and jet-novel subjet~\cite{Apolinario:2017qay} put new insight into the energy-loss process and, ultimately, the properties of the QGP itself.
 The modifications of jet fragmentation functions~\cite{Chatrchyan:2014ava,Aaboud:2017bzv,Sirunyan:2018qec,Aaboud:2019oac} as well as jet shape~\cite{Chatrchyan:2013kwa,Sirunyan:2018ncy} show that larger fraction of jet energy is carried by soft particles at large distances from the jet axis in Pb+Pb collisions due to the strong interactions between jet partons and medium constituents.  Angular distributions of charged particles around the jet axis in Pb+Pb and p+p collisions are also studied~\cite{CMS:2018zze,CMS:2016qnj,Aad:2019igg,Luo:2021hoo}, which indicate that yields of charged particles with low transverse momenta are observed to be increasingly enhanced, while charged particles with high transverse momenta have a suppressed yield in Pb+Pb collisions at larger distances.  However, those jet substructures are related to the particle number density which is rather sensitive to  hadronization  processes and  underlying background contamination of soft particles.  Recently, jet spectra measurements are extended to large area jets~\cite{CMS:2021vui}, with an anti-$k_t$ distance parameter R up to 1.0. The new data place
further constraints on the underlying jet quenching mechanisms, as well as a challenge to theoretical models with considerations of  hadronization effects and large background.  Besides, the grooming technique has been applied in~\cite{Sirunyan:2017bsd,Sirunyan:2018gct,Acharya:2019djg,Andrews:2018jcm,Casalderrey-Solana:2019ubu} by removing soft
wide-angle radiation from the jet, and so far focuses on the small radius jet.

In this paper we studied jets in Pb+Pb
collisions using a large radius, R = 1.0 with a reclustering procedure adopted by the recent ATLAS measurements~\cite{ATLAS:2019rmd}. The R = 1.0 jets are clustered from small-radius, R = 0.2, anti-$k_t$
jets with $p_T >$ 35 GeV/$c$ reconstructed using the standard procedure  described in Sec.~\ref{numerical}.  This procedure limits the impact of the underlying event but also does not allow recovering the energy transferred outside the R = 0.2 subjets via jet quenching.
The $k_t$ jet finding algorithm is used to re-cluster R = 1.0 jet constituents to obtain the $k_t$ splitting scales. Therefore, this measurement studies the hardest parton splittings and is not so sensitive to the hadronization processes.

In this paper, a Monte Carlo event generator PYTHIA8~\cite{Sjostrand:2014zea} is employed to simulate initial hard partons with shower partons, and the framework of Linear Boltzmann Transport (LBT) Monte Carlo model \cite{Li:2010ts,Wang:2013cia,He:2015pra,Cao:2016gvr,Cao:2017hhk} is used to study  the interaction and propagation of those partons in hot/dense QGP medium. With this framework, medium  modifications on the large-radius jet production as well as its hardest parton splitting  scale in Pb+Pb collisions relative to that in p+p collisions are
investigated.

The remainder of the paper is organized as follows. After providing a brief introduction to the LBT model in Sec. \ref{LBT},
we show in Sec. \ref{numerical} results from LBT on the large-radius jets as well as its hardest splitting.    We present the averaged jet energy loss responsible for the nuclear modification factor for large-radius and inclusive jets. The splitting scale $\sqrt {d_{12}}$  characterizing the hardest splitting of the large radius jets is investigated in the following in Sec.~\ref{numerical}.
The $R_{AA}$ sharply goes down with increasing $\sqrt{d_{12}}$ for small values of the splitting scale followed by flattening for larger $\sqrt{d_{12}}$. The medium modifications  of $\sqrt{d_{12}}$  distributions result from both the reduction of jet yields and the alteration of jet substructure. We also demonstrate that large radius jets with small splitting angle and small  fragmentation function are less suppressed.  A summary and some discussions are given in Sec.~\ref{summary}.


\section{The Linear Boltzmann Transport model}
\label{LBT}

A Linear Boltzmann Transport (LBT) model is employed to consider both elastic and inelastic scattering processes of both the initial jet shower partons and the thermal recoil partons with the quasi-particle in the QGP medium~\cite{Wang:2013cia, He:2015pra, Cao:2016gvr}. The elastic scattering process is simulated by the linear Boltzmann transport equation,
\begin{eqnarray}
&p_1\cdot\partial f_a(p_1)=-\int\frac{d^3p_2}{(2\pi)^32E_2}\int\frac{d^3p_3}{(2\pi)^32E_3}\int\frac{d^3p_4}{(2\pi)^32E_4} \nonumber \\
&\frac{1}{2}\sum _{b(c,d)}[f_a(p_1)f_b(p_2)-f_c(p_3)f_d(p_4)]|M_{ab\rightarrow cd}|^2 \nonumber \\
&\times S_2(s,t,u)(2\pi)^4\delta^4(p_1+p_2-p_3-p_4)
 \end{eqnarray}
where $f_{i=a,b,c,d}$ are the phase-space distributions of jet shower partons, $|M_{ab\rightarrow cd}|$ are the corresponding elastic matrix elements which are regulated by a Lorentz-invariant regulation condition $S_2(s,t,u)=\theta(s>2\mu^{2}_{D})\theta(-s+\mu^{2}_{D}\leq t \leq -\mu^{2}_{D})$. $\mu_{D}^{2}=g^{2}T^{2}(N_{c}+N_{f}/2)/3$ is the Debye screening mass. The inelastic scattering is described by the higher twist formalism for induced gluon radiation as~\cite{Guo:2000nz,Zhang:2003yn,Zhang:2004qm,Zhang:2003wk},
\begin{equation}
\frac{dN_g}{dxdk_\perp^2 dt}=\frac{2\alpha_sC_AP(x)\hat{q}}{\pi k_\perp^4}\left(\frac{k_\perp^2}{k_\perp^2+x^2M^2}\right)^2\sin^2\left(\frac{t-t_i}{2\tau_f}\right)  .
 \end{equation}
Here $x$ denotes the energy fraction of the radiated gluon relative to parent parton with mass $M$, $k_\perp$ is the transverse momentum. A lower energy cut-off $x_{min}=\mu_{D}/E$ is applied for the emitted gluon in the calculation. $P(x)$ is the splitting function in vacuum, $\tau_f=2Ex(1-x)/(k^2_\perp+x^2M^2)$ is the formation time of the radiated gluons in QGP. The dynamic evolution of bulk medium is given by 3+1D CLVisc hydrodynamical model~\cite{Pang:2012he, Pang:2014ipa} with initial conditions simulated from A Multi-Phase Transport (AMPT) model~\cite{Lin:2004en}. Parameters used in the CLVisc are fixed by reproducing hadron spectra with experimental measurements. In LBT model, $\alpha_s$ is the strong coupling constant which is served as the only one parameter to control the strength of parton-medium interaction. Based on the previous studies~\cite{He:2018xjv}, we choose $\alpha_s=0.15$ for the following calculations. LBT model has been well tested that could provide nice description of a series of jet quenching measurements, from light and heavy flavor hadrons suppression to single inclusive jets suppression, as well as bosons-jet correlations~\cite{He:2015pra, Cao:2016gvr, Luo:2018pto, Zhang:2018urd}.

\section{Results and Analysis}
\label{numerical}
 In our calculations, initial hard partons with shower partons are generated by PYTHIA8~\cite{Sjostrand:2014zea}, with the infrared (IR) cut-off  1 GeV/$c$, and then those partons will propagate through the hot-dense medium within LBT~\cite{Wang:2013cia, He:2015pra, Cao:2016gvr}. In order to compare with the experimental data, the jet reconstruction procedures~\cite{ATLAS:2019rmd}  and the underlying event subtraction are performed following the ATLAS experiment~\cite{ATLAS:2019rmd,Aad:2012vca}.
 The small-radius jets are reconstructed using the anti-$k_t$  algorithm~\cite{Cacciari:2008gp} implemented in the FastJet~\cite{Cacciari:2011ma} software package with  $R$ = 0.2. Those small-radius jets  are  accepted in the rapidity range of $|y|<3.0$ and in transverse momentum range of $p_T^{jet}>35$ GeV/$c$.
 The anti-$k_t$ algorithm with distance parameter $R$ = 1 is used to recluster the small-radius, $R=0.2$ jets. These resulting jets are then referred to as reclustered large-radius (LR) jets in the following.  In the calculation we impose kinematic cut that $|y^{jet}|<2.0$.

\begin{figure}
  \centering
\includegraphics[scale=0.34]{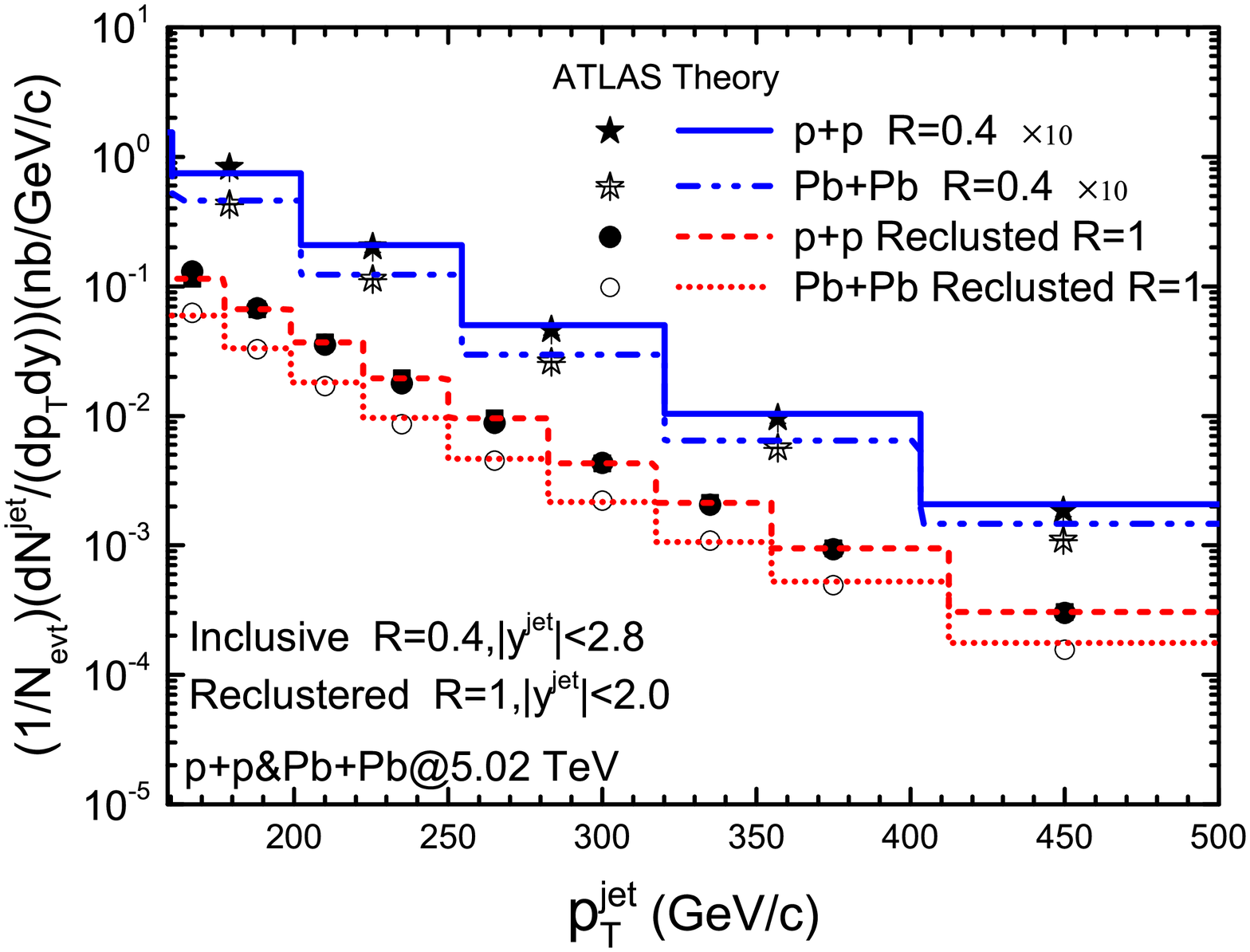}\vspace{-20pt}
\includegraphics[scale=0.34]{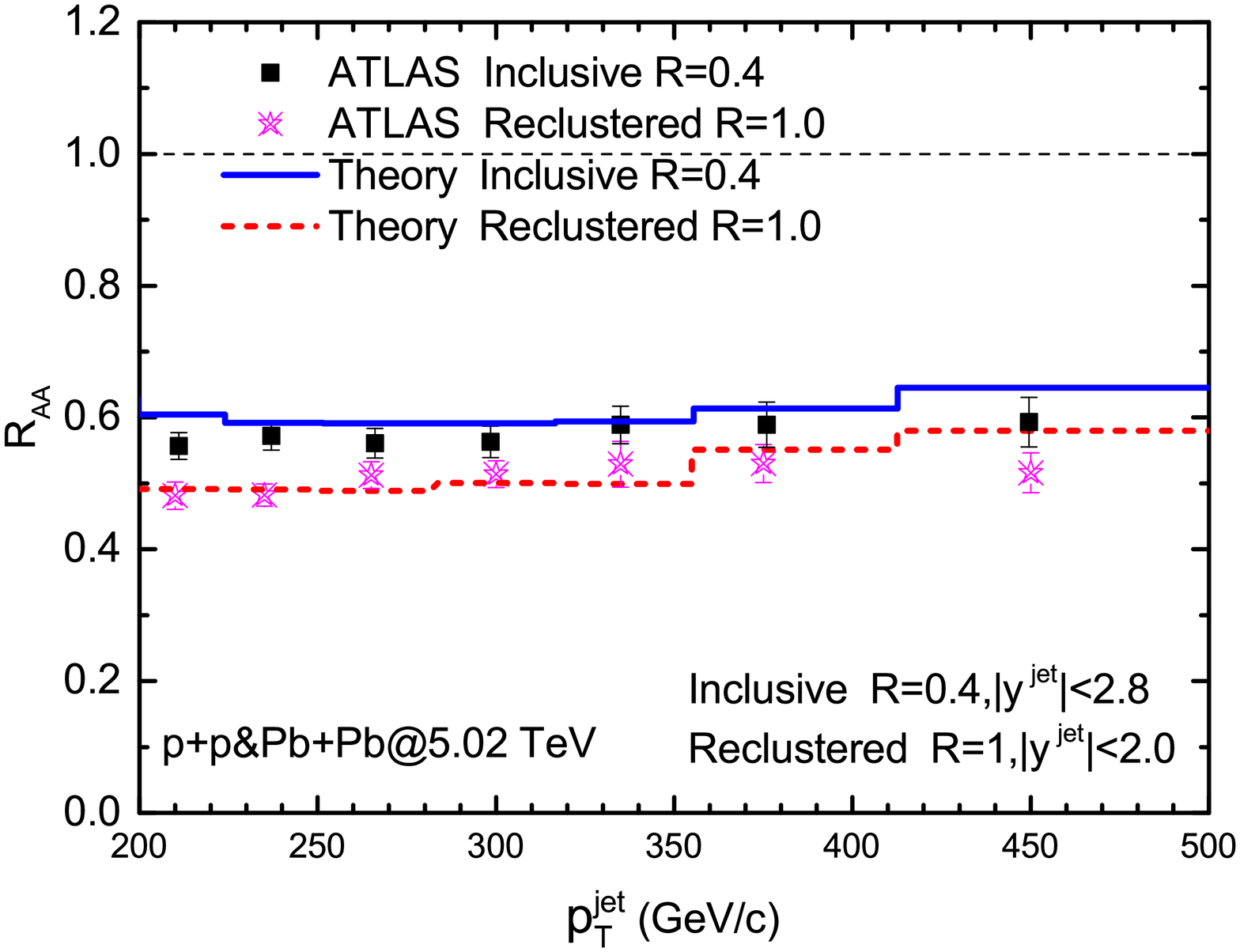}
  \caption{(Color online) Distributions of reclustered  LR jets  and  inclusive jets in both  p+p and Pb+Pb collisions with the comparison against the experimental data~\cite{ATLAS:2019rmd,Aaboud:2018twu} (top).   Nuclear modification factor $R_{AA}(p_T^{jet})$ of  reclustered LR jets  and  inclusive jets with comparison against the experimental data~\cite{ATLAS:2019rmd,Aaboud:2018twu} (bottom).   }\label{pt_jet}
\end{figure}

The distributions of reclustered LR jets as a function of jet transverse momentum $p_T^{jet}$ in both p+p and Pb+Pb collisions are calculated at $\sqrt{s_{NN}}=5.02$ TeV and compared to experimental data~\cite{ATLAS:2019rmd}, as presented in the top panel of Fig.~\ref{pt_jet}. Distributions of  the  inclusive $R=0.4$ jet are also plotted with the corresponding experimental data~\cite{Aaboud:2018twu}.  Our numerical calculations show well agreements with the experimental data  for both inclusive jets and reclustered LR jets in both p+p and Pb+Pb collisions. An obvious spectra suppression of the inclusive jet  as well as the reclustered LR jet in Pb+Pb collisions relative to p+p collisions can be seen.  The nuclear modification factors of those spectra, $R_{AA}$, evaluated as a function of jet transverse momentum $p_T^{jet}$ are compared to experimental data~\cite{ATLAS:2019rmd,Aaboud:2018twu} in the bottom panel of Fig.~\ref{pt_jet}.  Our model calculations give nice descriptions of the experimental data,
and show that  $R_{AA}$ goes up slowly with $p_T^{jet}$.  It is interesting to observe that the nuclear modification factor for reclustered LR jets with $R=1.0$ is a bit smaller than the one of inclusive  jets with $R=0.4$, which results from the fact that with its specific jet-finding algorithm reclustered LR jets exhibit quite different structures from inclusive jets with conventional definition.

\begin{figure}
  \centering
  \includegraphics[scale=0.34]{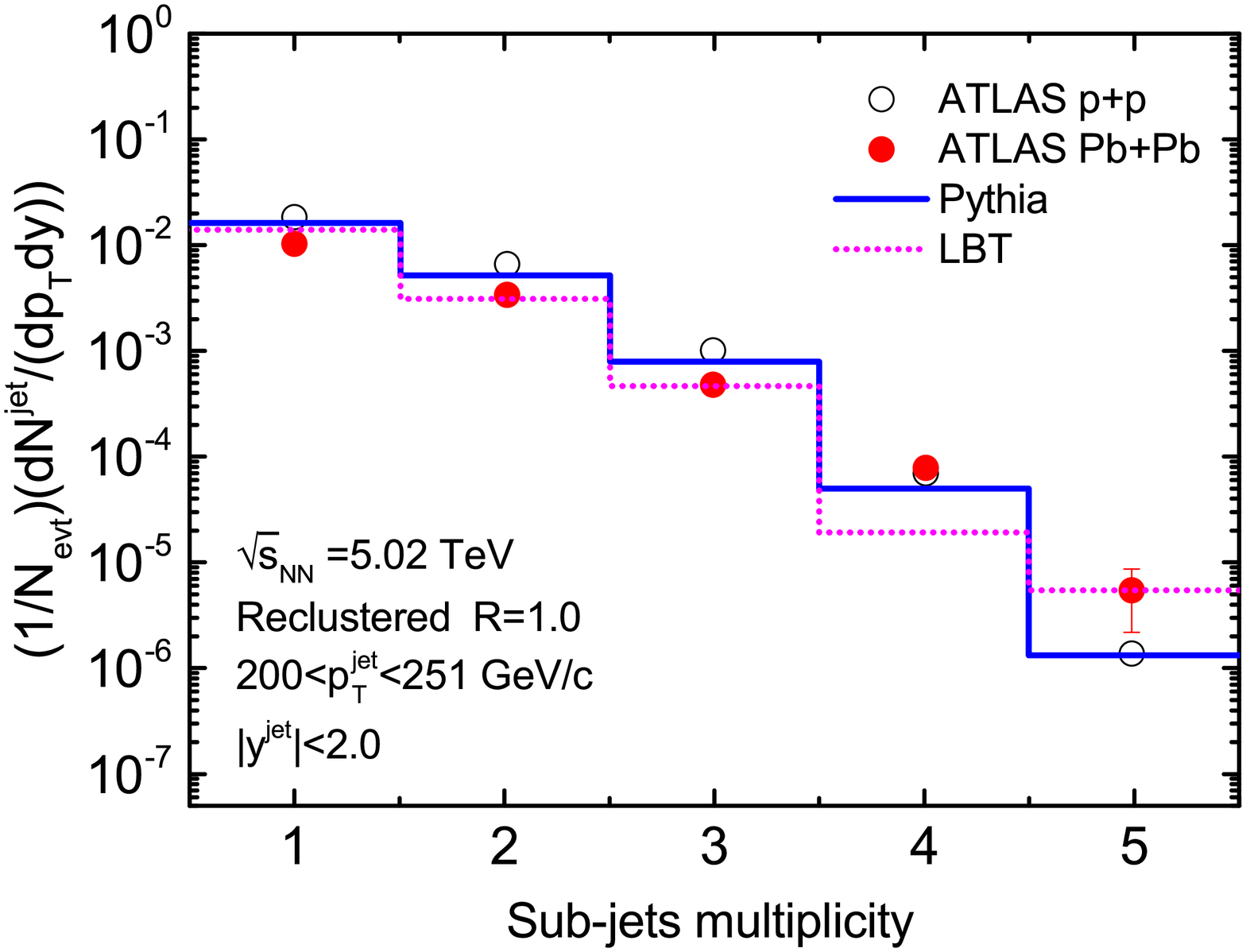}\vspace{-20pt}
  \includegraphics[scale=0.34]{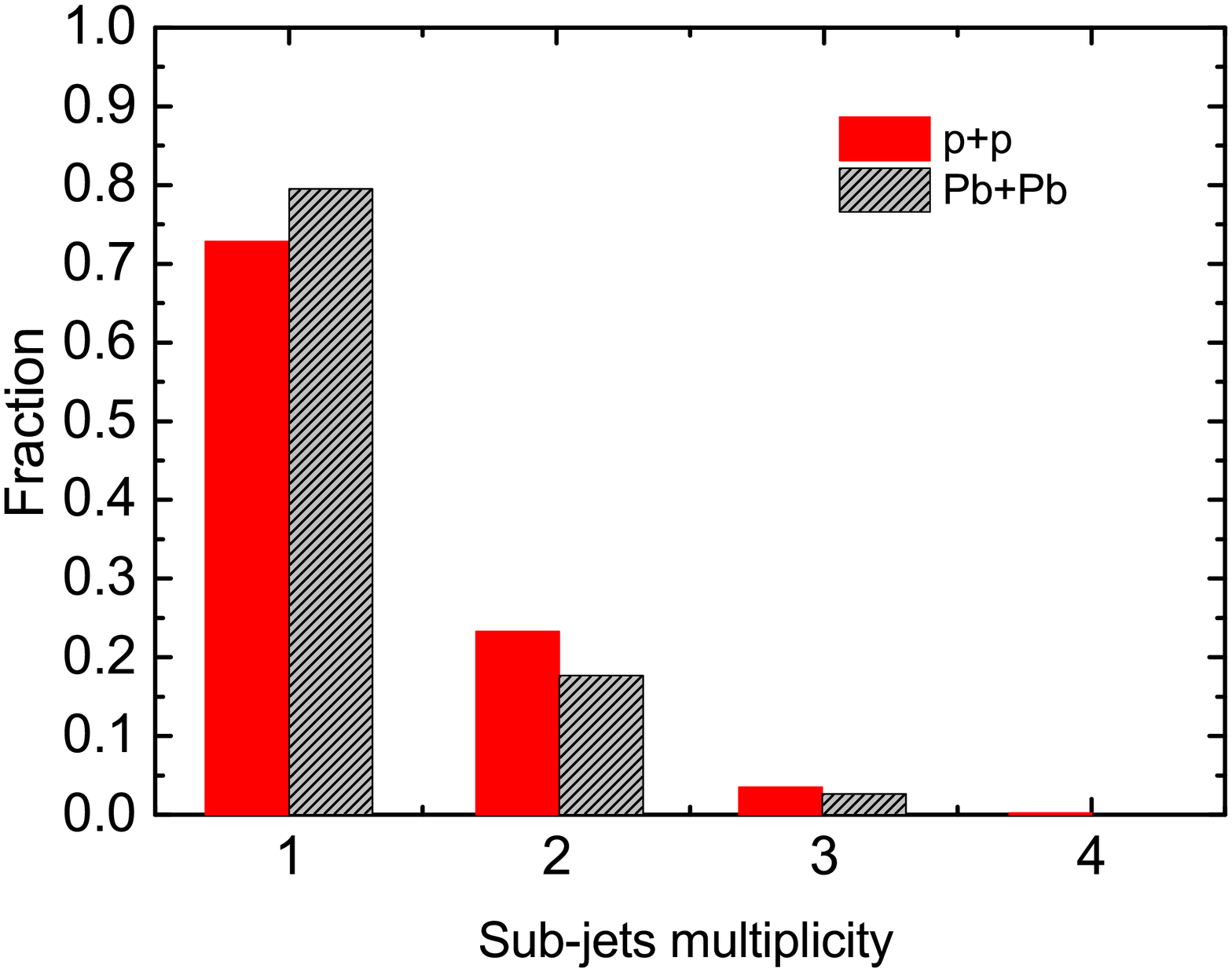}
  \caption{(Color online)  Number of the subjet of  reclustered  LR jets before and after jet quenching and the comparison with ATLAS preliminary data (top panel). Fraction of  N-subjet of reclustered  LR jet in Pb+Pb and p+p collisions (bottom panel).  }\label{Nsubjet}
\end{figure}

\begin{figure}
  \centering
\includegraphics[scale=0.34]{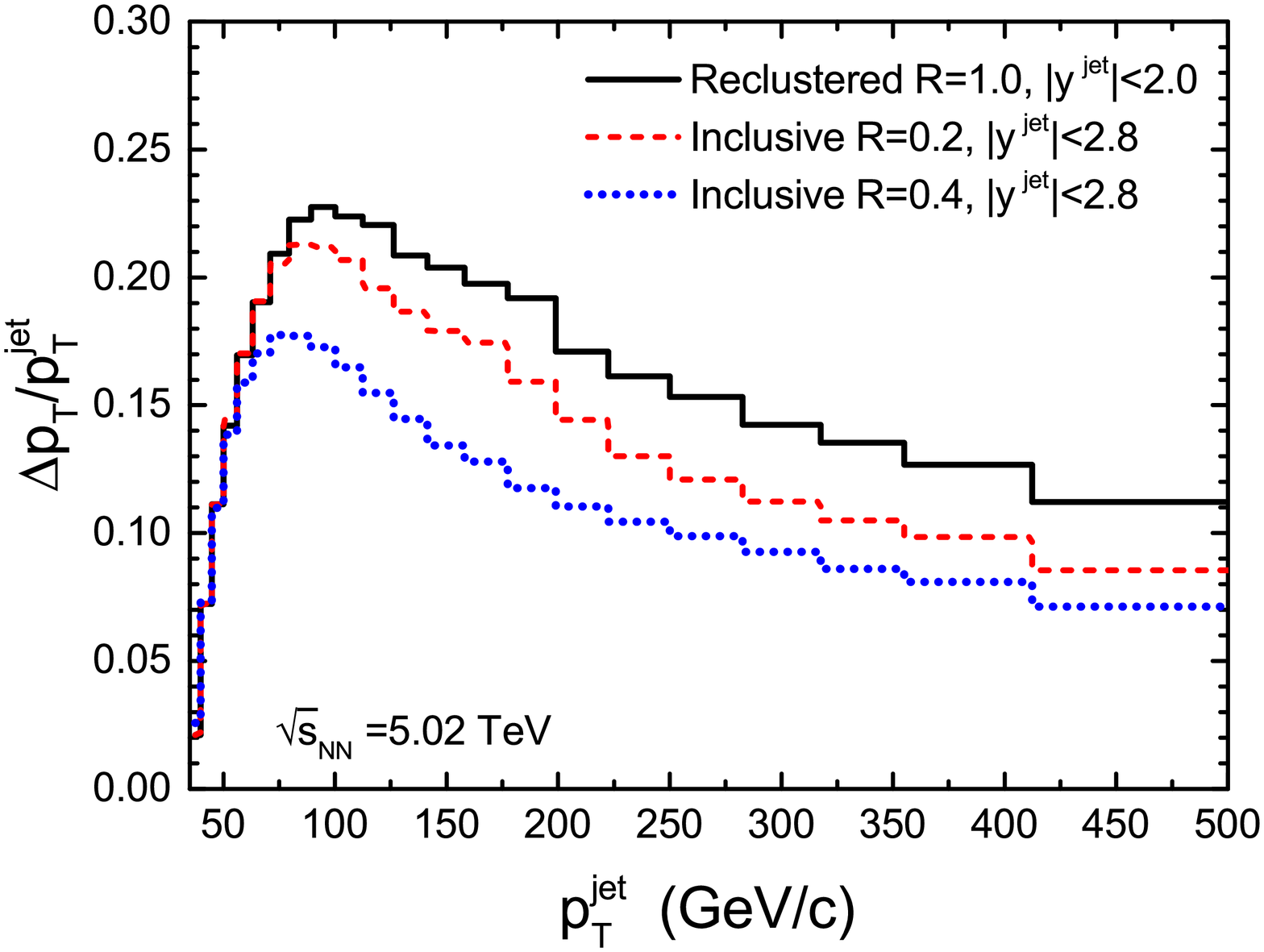}\vspace{-20pt}
 \includegraphics[scale=0.34]{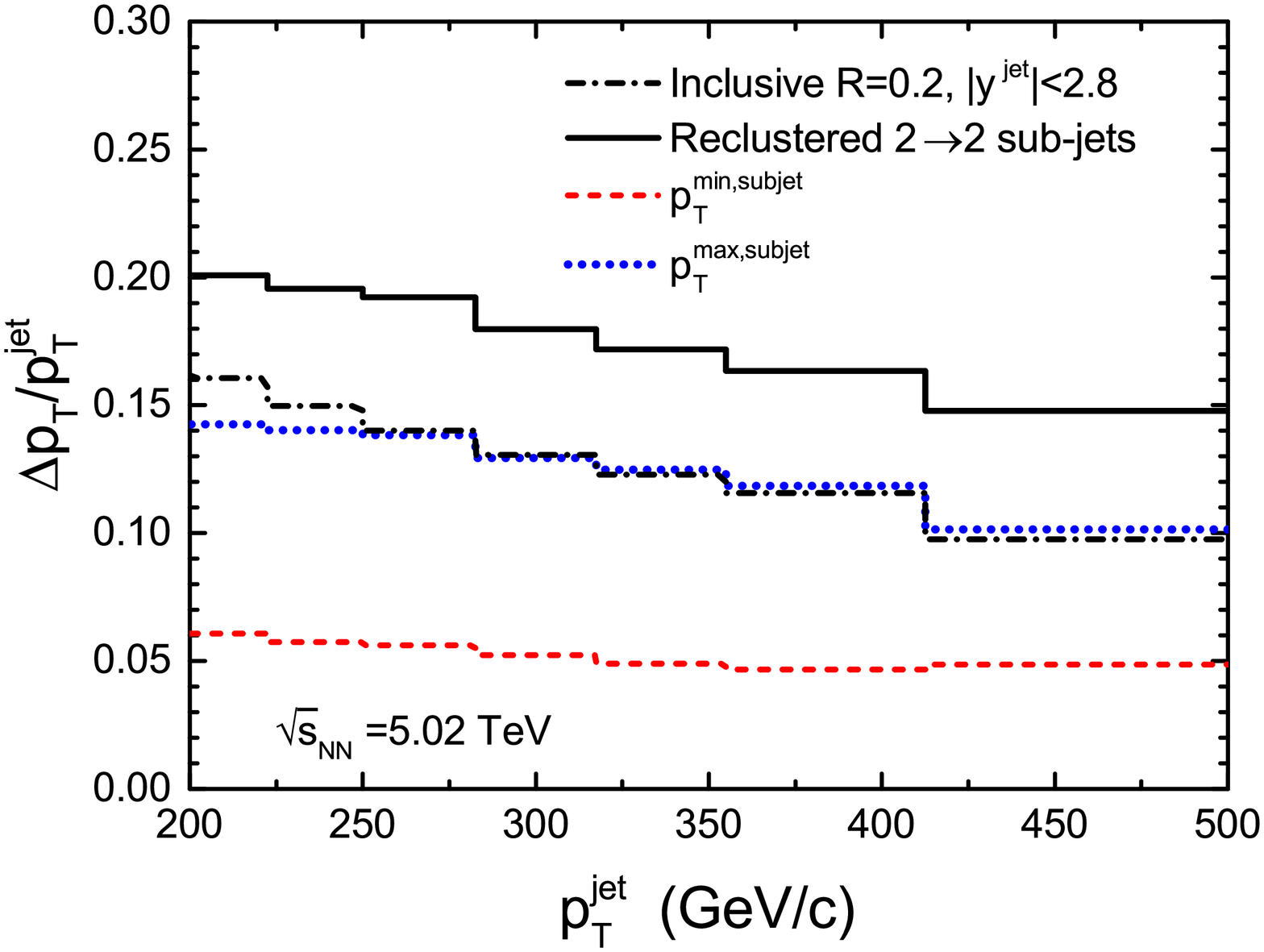}
  \caption{(Color online)  Fraction transverse momentum loss of reclustered LR  jets  as well as  inclusive jets  as a function of jet transverse momentum in vacuum (top panel).  Ratio of  transverse momentum loss of reclusteed LR  jets with two subjets both before and after jet quenching as well as its leading subjet and subleading constituent  to  initial LR jet $p_T$  evaluated as a function of the vacuum $p_T^{jet}$ at $\sqrt{s_{NN}}=5.02$ TeV (bottom panel).  }\label{average_energy_loss}
\end{figure}

We plot the fraction of the subjet multiplicity of LR jet in both p+p and Pb+Pb collisions in Fig.~\ref{Nsubjet}. As can be seen, almost 70$\%$ of reclustered LR jets have only one subjet, and in this case these reclustered LR jets are just  inclusive jets with $R=0.2$. Besides, compared to p+p collisions, the fraction of the LR jet with only a single subjet is a bit increased, while the fraction of large-radius jets with complex substructures  is decreased.
Those modifications indicate that jets with one subjet may lose on average less energy than jets with multiple subjets, and the change of jet substructure due to jet-medium interactions may also play a role, which will be discussed in more details later.

To illustrate the transverse momentum and  jet cone dependence of jet energy loss, we calculated the average $p_T$ reduction of reclustered LR jets with  $R=1$  as well as  inclusive jets with $R=0.2$ and $R=0.4$  in the medium as a function of jet transverse momentum $p_T^{jet}$, shown in Fig.~\ref{average_energy_loss}. In the calculations,  partons  within a distance $R=2.0$ of the reconstructed jet in each
event are identified. These partons are then used for the re-construction of the vacuum leading jet in p+p collisions
with a given jet-cone size R. They propagate through the hot-dense medium through LBT and then are collected as the medium modified jets with the same jet algorithm.

The comparison between inclusive $R=0.2$ and $R=0.4$ jets in Fig.~\ref{average_energy_loss} shows that  the averaged transverse momentum loss is significantly reduced for inclusive jets with large jet cone as expected
in~\cite{Vitev:2008rz,Vitev:2009rd,He:2011pd,Chien:2015hda,Aad:2012vca}. While  the averaged transverse momentum attenuation of reclustered LR jet with $R=1.0$ is a little larger than that of the inclusive jet with $R=0.2$.

 The analysis in Fig.~\ref{average_energy_loss} shows that the fraction first increases with $p_T^{jet}$ in the region $p_T^{jet}<80$ GeV/$c$, as also illustrated in Refs.~\cite{Zhang:2018urd,Zhang:2018kjl,Zhang:2021oki}, and reaches its peak value (about 20$\%$) near $p_T^{jet}\simeq 80$~GeV/$c$, then goes down with $p_T^{jet}$ in the region $p_T^{jet}>80$ GeV/$c$.  The downtrend in the large $p_T^{jet}$ region leads to slow increasing nuclear modification factor with $p_T^{jet}$.

We notice that the reclustered LR jet with $R=1$ containing only one subjet is just the same with inclusive $R=0.2$ jet.  However, these reclustered  LR jets with two subjets both before and after jet quenching (denoted as  ``Reconstructed $ 2 \rightarrow 2$ subjets " ), will lose much larger fraction of their energies than inclusive $R=0.2$ jets.  The ratios of average energy loss $\Delta p_T$  of the reclustered LR jet with two subjets both before and after jet quenching, as well as the leading subjet and subleading constituent to the initial jet energy $p_T^{jet}$ before quenching  are shown in the bottom panel of Fig.~\ref{average_energy_loss}. Though the leading subjets have relative smaller energy compared to inclusive $R=0.2$ jets, those small radius jets may lose relative larger fraction of its energy compared to inclusive $R=0.2$ jets with larger transverse momentum. Therefore we see that the leading subjets lose almost the same amount of energy as the inclusive $R=0.2$ jets.  Besides, the subleading constituent will also lose some amount of energy.  Consequently,  the energy loss of the reclustered LR jets should be the sum of the individual subjets, and is found to be larger than that  of inclusive $R=0.2$ jets at fixed $p_T^{jet}$.  As a result, the nuclear modification factor for the reclustered LR jets with $R=1.0$ is a bit smaller than that for inclusive $R=0.2$ jets.

\begin{figure}
  \centering
\includegraphics[scale=0.34]{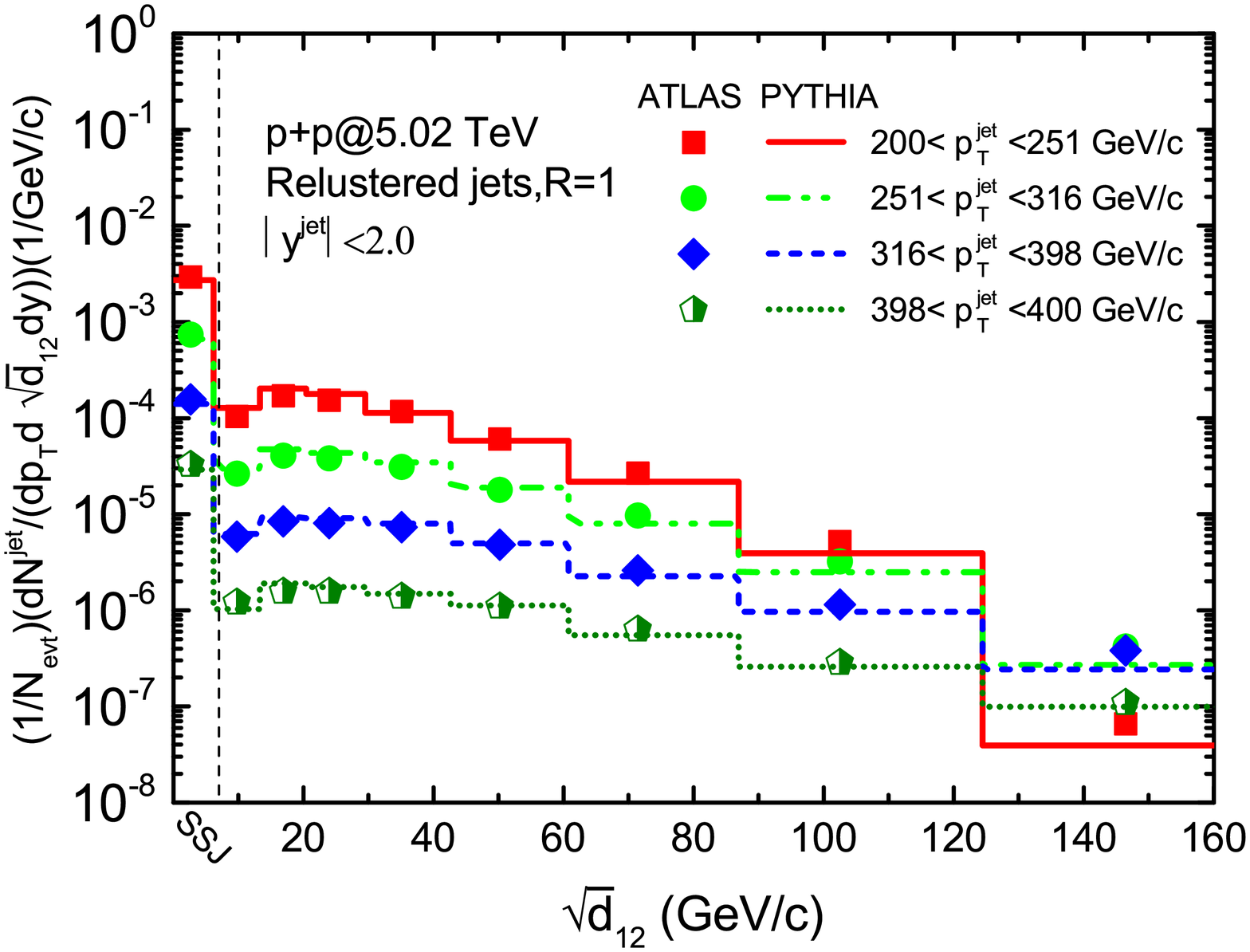}\vspace{-20pt}
\includegraphics[scale=0.34]{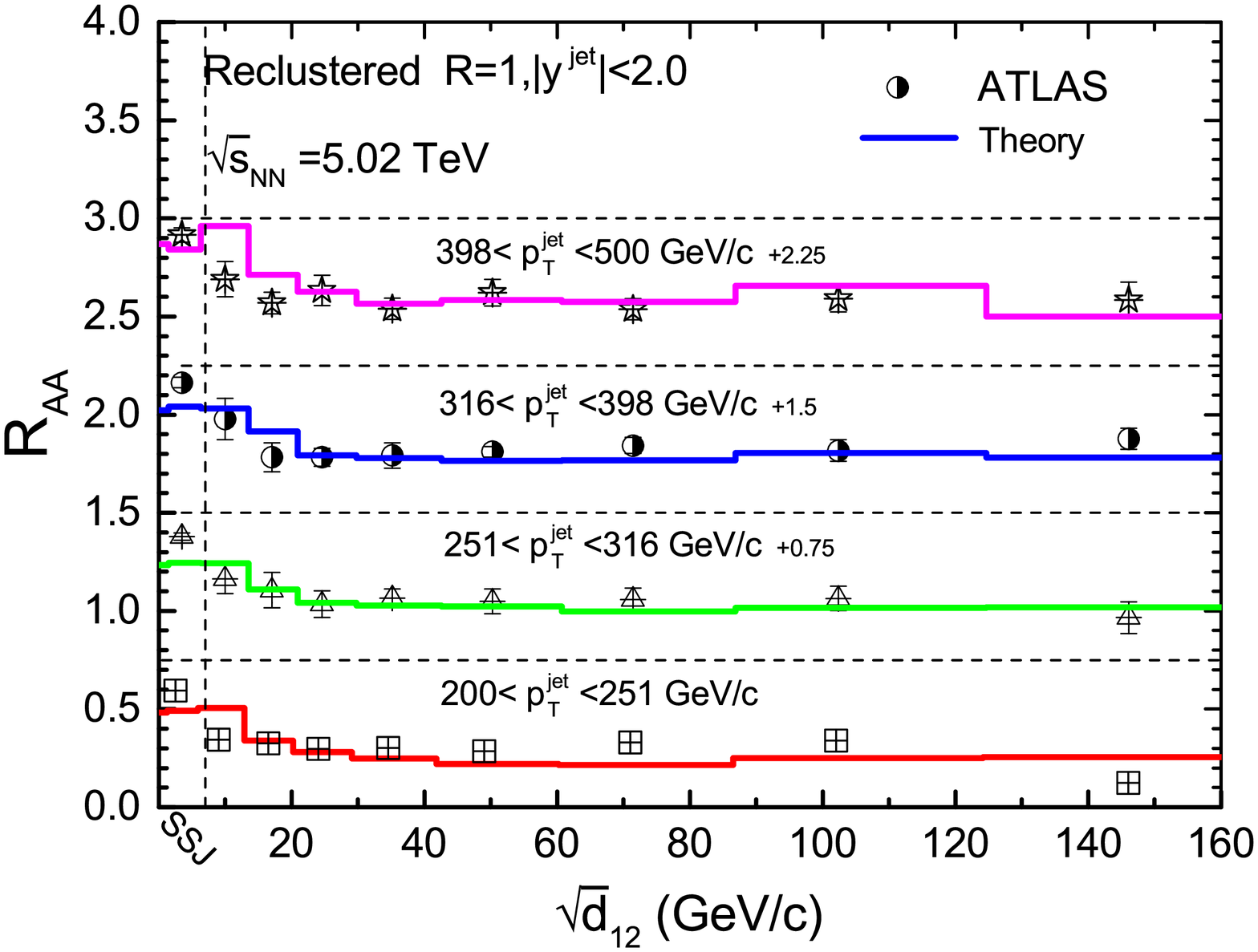}
  \caption{(Color online) Distributions of splitting scale in different reclustered LR jet transverse momentum intervals in p+p collisions as well as the comparison with the experimental data~\cite{ATLAS:2019rmd} (top panel).  Nuclear modification factor  evaluated as a function of  the splitting scale $\sqrt{ d_{12}}$ in different reclustered LR jet transverse momentum intervals as well as the comparison with the experimental data~\cite{ATLAS:2019rmd} at $\sqrt{s_{NN}}=5.02$ TeV (bottom panel).   }\label{raa_d12}
\end{figure}



To obtain a deeper understanding of the medium modification on jet fragmentation pattern due to jet-medium interactions, we calculated the medium modification on the hardest jet  splitting processes.  We recluster the large-R jet constituents  with $k_t$  jet algorithm~\cite{Ellis:1993tq} which combines the most soft or the closest parton firstly. Therefore, the last step of the combining procedure  should be the hardest splitting process.
The splitting scales of the hardest splitting process are determined
by clustering objects together according to their distance from each other with the following distance definition:
\begin{eqnarray}
 &&d_{12}=min(p_{T1}^2,p_{T2}^2)\cdot  \Delta R_{12}^2/R^2,\ \ \ \  \nonumber \\
 &&\Delta R_{12}=\sqrt{\Delta\phi_{12}^2+\Delta y_{12}^2}
\end{eqnarray}
where $p_{T1}$, $p_{T2}$  are the constituent transverse momentum of the hardest splitting,  $\Delta\phi_{12}$, $\Delta y_{12}$, $ \Delta R_{12}$ are the azimuthal angle, rapidity and distance between the two  splitting constituents respectively. For reclustered LR jets with only one single subjet (SSJ), $\sqrt{d_{12}}$ is  equal to zero. In this work,  distribution of $\sqrt{d_{12}}=0$  is shown in the bin $0<\sqrt{d_{12}}<7$, and donated as ``SSJ" in Fig.~\ref{raa_d12}.
The distributions of reclustered LR jets as a function of the splitting scale $\sqrt{d_{12}}$ are calculated in four $p_T^{jet}$ intervals in p+p collisions and compared to experimental data~\cite{ATLAS:2019rmd} in the top panel of Fig.~\ref{raa_d12}. Our numerical results show well agreement with the experimental data in all $p_T^{jet}$ intervals.


The nuclear modification factors evaluated as a function of $\sqrt{ d_{12}}$  in four $p_T^{jet}$ intervals are  also calculated  and compared to the experimental data in the bottom panel of Fig.~\ref{raa_d12}. Our simulation results can provide well description to  the experimental  measurements on the suppression of the splitting scale $\sqrt{ d_{12}}$. The distributions  are all suppressed in Pb+Pb collisions compared to p+p collisions in all $p_T^{jet}$ intervals. The $R_{AA}$ sharply decreases with increasing $\sqrt{d_{12}}$ for small values of the splitting scale followed by flattening for larger $\sqrt{d_{12}}$.

\begin{figure}
\centering
\includegraphics[width=0.54\textwidth]{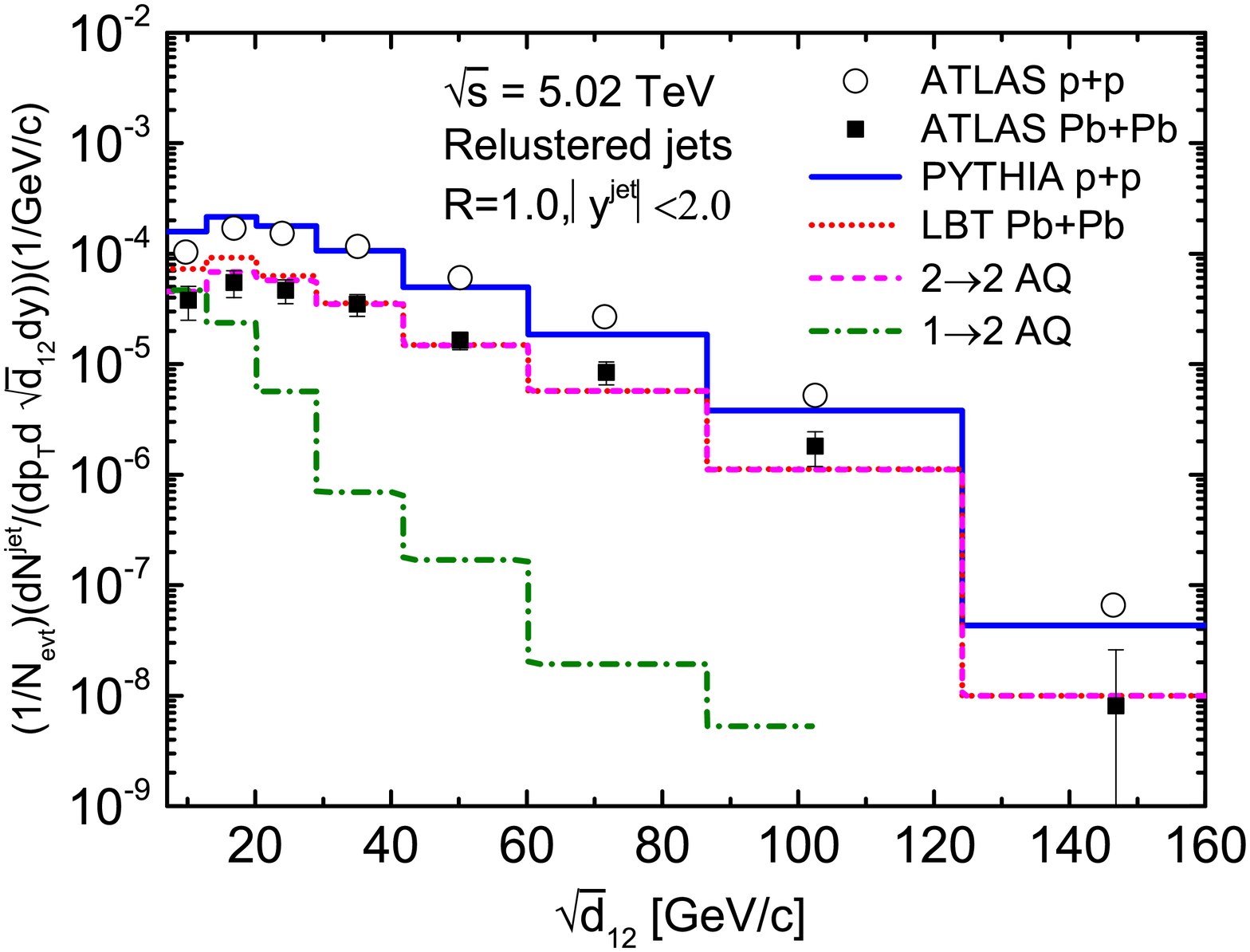}\vspace{-20pt}
\includegraphics[width=0.54\textwidth]{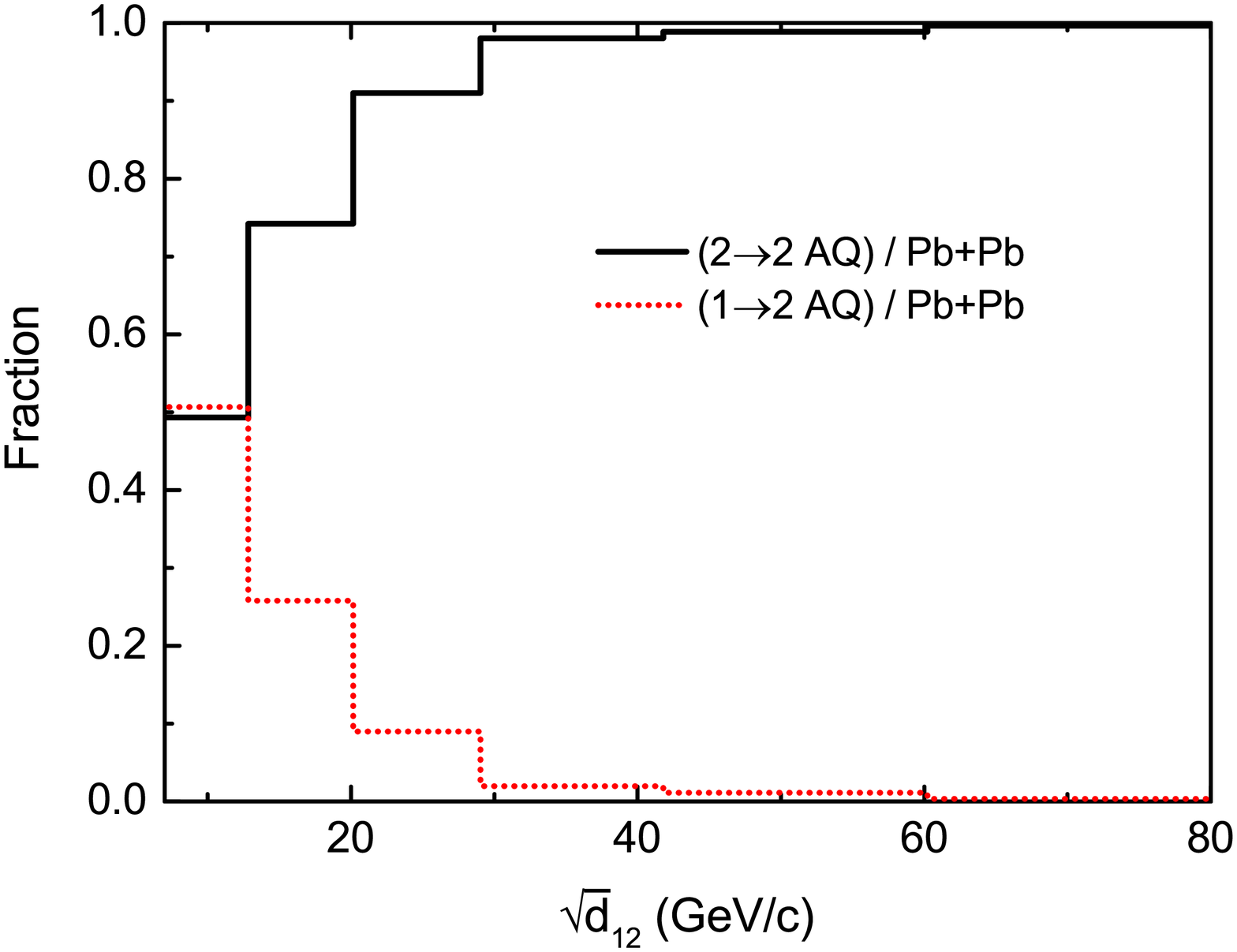}
\vspace{-8pt}
  \caption{(Color online) Distributions of splitting scale $\sqrt{d_{12}}$ with $200<p_T^{jet}<251~$GeV/c in p+p and Pb+Pb collisions as well as the contributions from different processes in Pb+Pb (top), and the comparison with experimental data~\cite{ATLAS:2019rmd}.  The relative contribution  fractions from the process $2 \rightarrow 2$ (without changes of jet substructures) and $1 \rightarrow 2$ (with changes of jet substructures) to  jet $\sqrt{d_{12}}$ spectrum in Pb+Pb (bottom). } \label{d12_fraction}
\end{figure}

We emphasize that the medium modifications  on jet $\sqrt{ d_{12}}$  spectra are the combined result of both jet yields reduction and  the changes of the jet substructures in Pb+Pb collisions.
The magnitude of parton energy loss in the hot QGP may significantly decrease the LR jet yields (for instance, see Fig.~\ref{Nsubjet}), which then leads to an overall suppression of  the nuclear modification  factor $R_{AA}(\sqrt{ d_{12}})$.
It is noted that the selection bias~\cite{Brewer:2021hmh,Connors:2017ptx,Renk:2012ve} plays an important role for the reduction of jet yields, for instance, the numerical analysis shows that for LR jets with $200<p_T^{PbPb}<251$~GeV/c in Pb+Pb at $5.02$~TeV,  more than $60\%$ of them originate from unquenched jets with the same momentum interval $200<p_T<251$~GeV/c, whereas less than $40\%$ of them come from unquenched jets with $p_T>251$~GeV/c.
Fig.~\ref{d12_fraction} presents the distributions of splitting scale $\sqrt{d_{12}}$ with $200<p_T^{jet}<251~$GeV/c in p+p and Pb+Pb collisions as well as the contribution fractions to $\sqrt{d_{12}}$ distributions from $2 \rightarrow 2$ process and $ 1 \rightarrow 2$ process in Pb+Pb collisions.  Here  $2 \rightarrow 2$ stands for the process which has $2$ subjets both {\it before} and {\it after} quenching, and gives leading contribution for those processes without the change of subjet numbers (or jet substructures), since $\geq 3$ subjets events are much rarer, whereas $1$ subjet events give $\sqrt{d_{12}} = 0$.  The process $1 \rightarrow 2$  denotes the process which has $1$ subjet  {\it before quenching} but $2$ subjets {\it after quenching}.  We observe in the bottom panel of Fig.~\ref{d12_fraction} that $2 \rightarrow 2$ process gives $ >90\%$ contributions in the region with $\sqrt{d_{12}} > 30$~GeV/c.
However it is noted that in small $\sqrt{d_{12}}$ region the contribution of $2 \rightarrow 2$ process has been reduced to
about $50\%$, and the modifications of jet substructure (for instance,  the process $ 1 \rightarrow 2$) will give significant contributions ( $\sim 50\%$ when $\sqrt{d_{12}}\sim 10$~GeV/c), which are indispensable in making a decent description of distributions of the splitting scale $\sqrt{d_{12}}$ in Pb+Pb collisions. Medium modifications on jet substructure may alter $\sqrt{d_{12}}$ distribution pattern, eg. $1\rightarrow 2$ process, increasing the yield of jet with small splitting scale, leads to less suppression in small splitting scale region,  while $ 2 \rightarrow 1$  process, reducing the yield with large splitting scale, leads to almost constant  suppression in large splitting scale region, as we may show below.

\begin{figure}
  \centering
\includegraphics[scale=0.34]{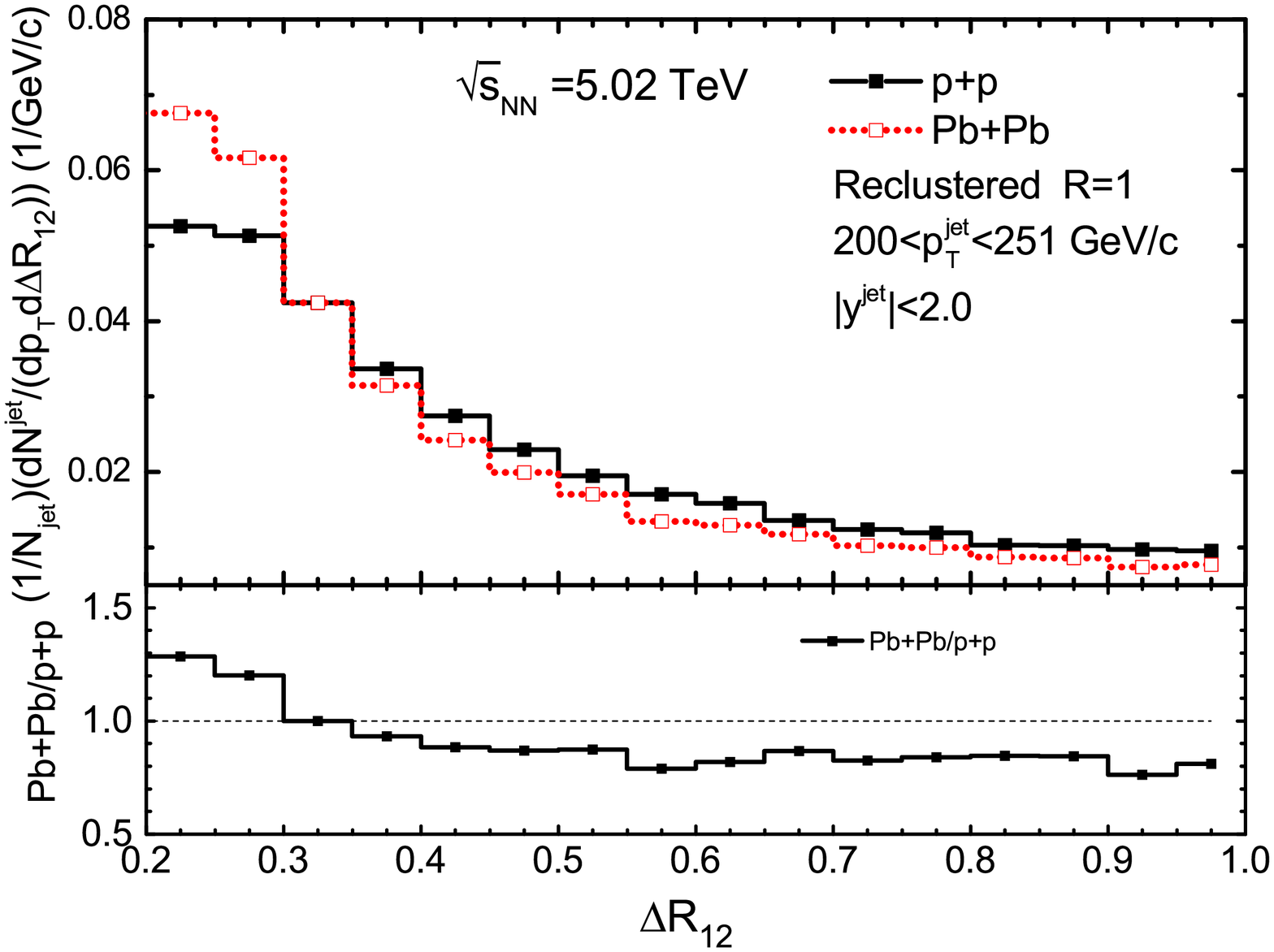}\vspace{-20pt}
\includegraphics[scale=0.34]{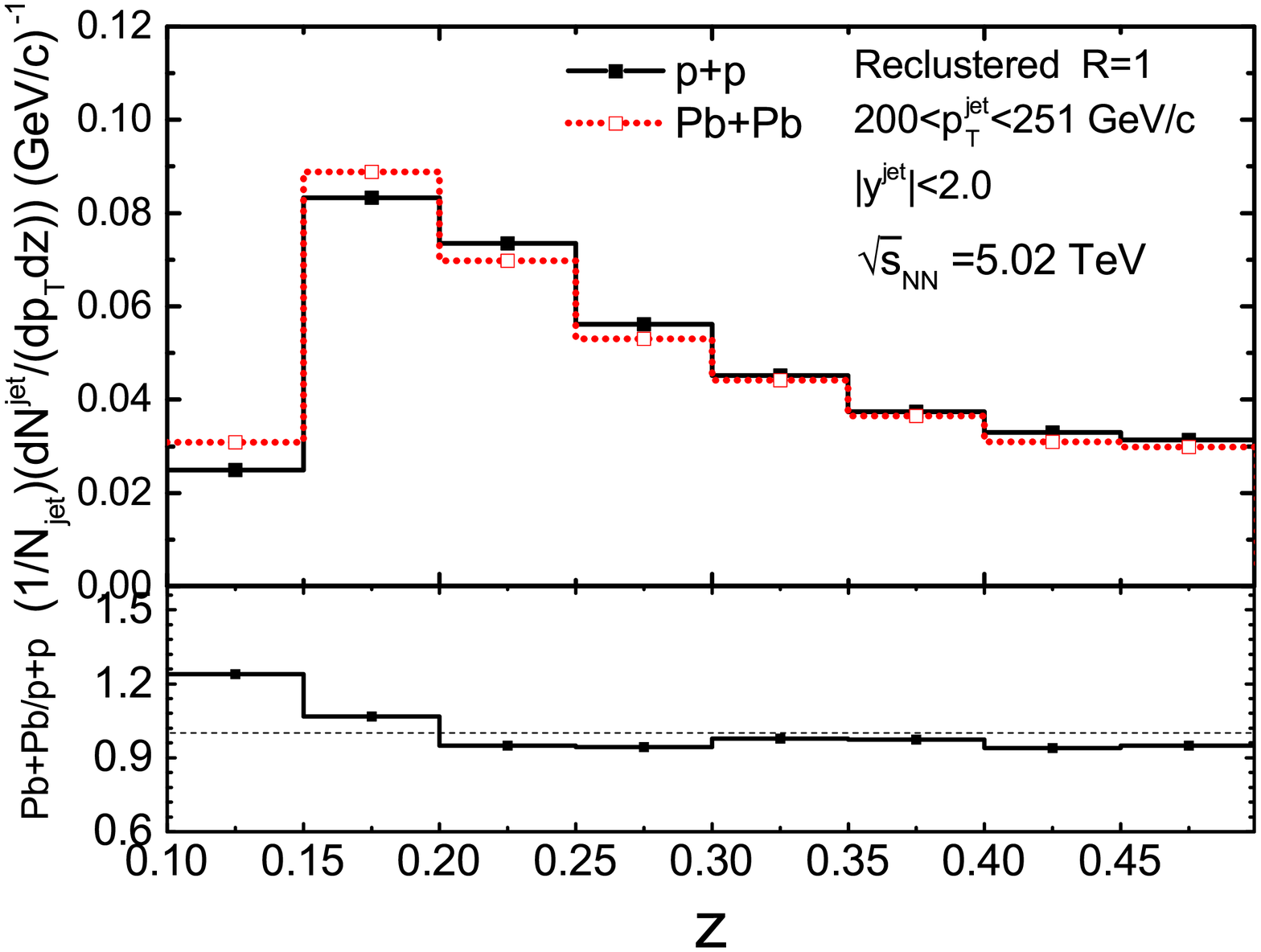}
  \caption{(Color online) Normalized LR jet distributions as a function of the  splitting scale $\Delta R_{12}$ in p+p and Pb+Pb collisions as well as its modification due to jet-medium interactions (top panel).    Normalized distributions  calculated as a function of the splitting variable $z=p_T^{min}/p_T^{jet}$ in both p+p and Pb+Pb collisions as well as its modification in Pb+Pb collisions relative to p+p collisions $\sqrt{s_{NN}}=5.02$ TeV (bottom panel). }\label{raa_r12_z}
\end{figure}

  According to the definition of the hardest splitting scale,  $\sqrt{d_{12}}$ is strongly correlated with the transverse momentum of soft-subjet $p_T^{Subjet}$(subjet with smaller energy) and $\Delta R_{12}$. To illustrate the $\sqrt{d_{12}}$ dependence suppression of  $R_{AA}$,  we calculated the modification of $ \Delta R_{12}$ and the fragmentation scale $z$  in Fig.~\ref{raa_r12_z}. The normalized distributions of $\Delta R_{12}$ and $z$ as well as their modification factors are presented in Fig.~\ref{raa_r12_z}.
  Both $\Delta R_{12}$ and $z$  will be 0 for  the large-radius jets with a single subjet.  In our calculations,  we only show the results of reclustered LR
jets of a complex substructure with more than one subjet, giving a non-zero $\sqrt{d_{12}}$.
We note that jets with distance $\Delta R_{12}<0.2$ may overlap with each other with a small chance, and could be considered as one jet in most of cases.  The kinematic threshold for the subjets is $p_T^{jet}> 35$ GeV/$c$, thus $z_{min}= (p_T^{Subjet})_{min}/(p_T^{jet})_{max}=35/250=0.14$. Besides, $p_T^{subjet}$ can hardly exceed half of the LR jets transverse momentum $p_T^{jet}$, so the value of $z$ is limited to the range 0.14-0.5.

\begin{figure}
  \centering
\includegraphics[scale=0.34]{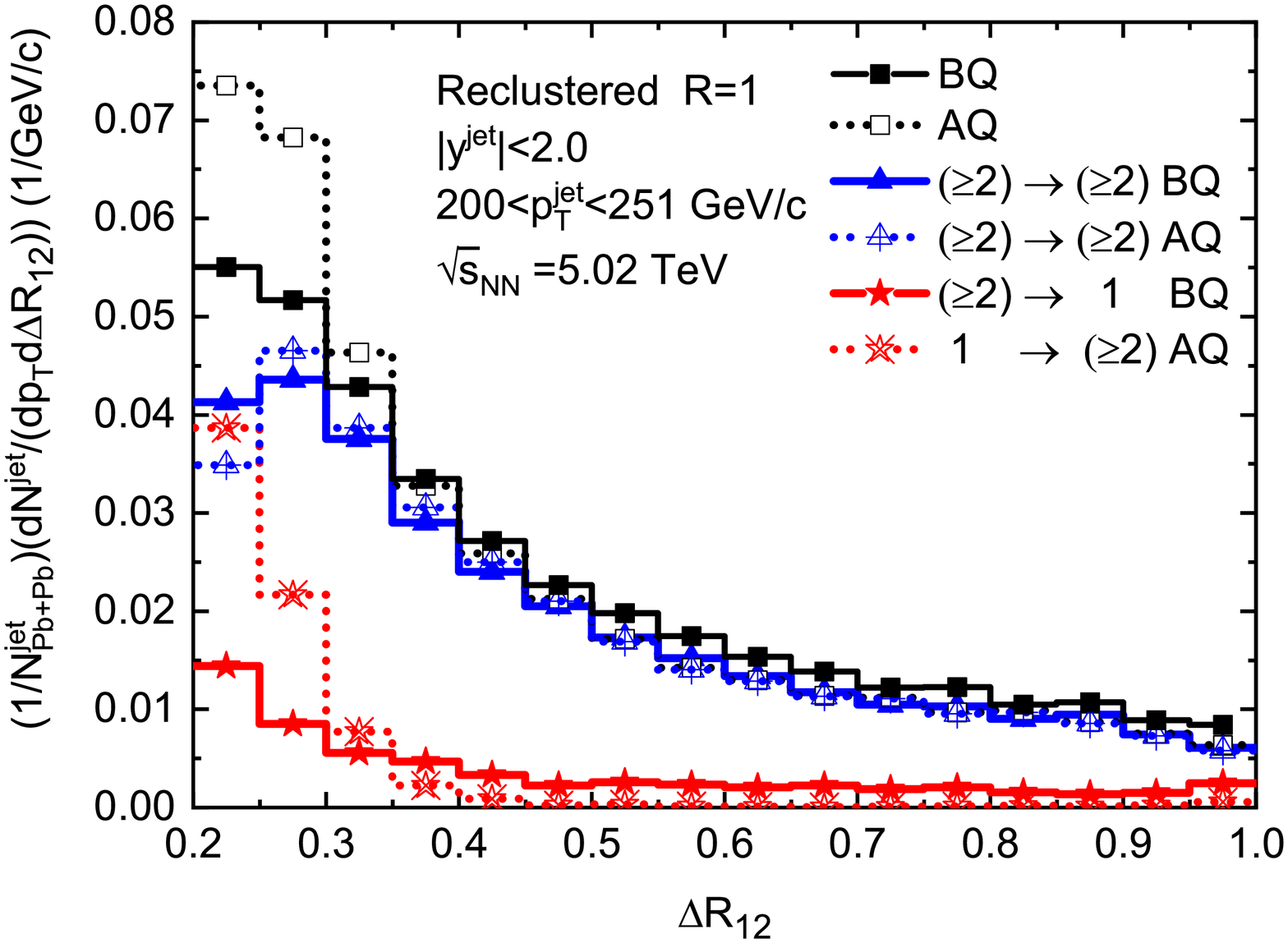}\vspace{-20pt}
\includegraphics[scale=0.33]{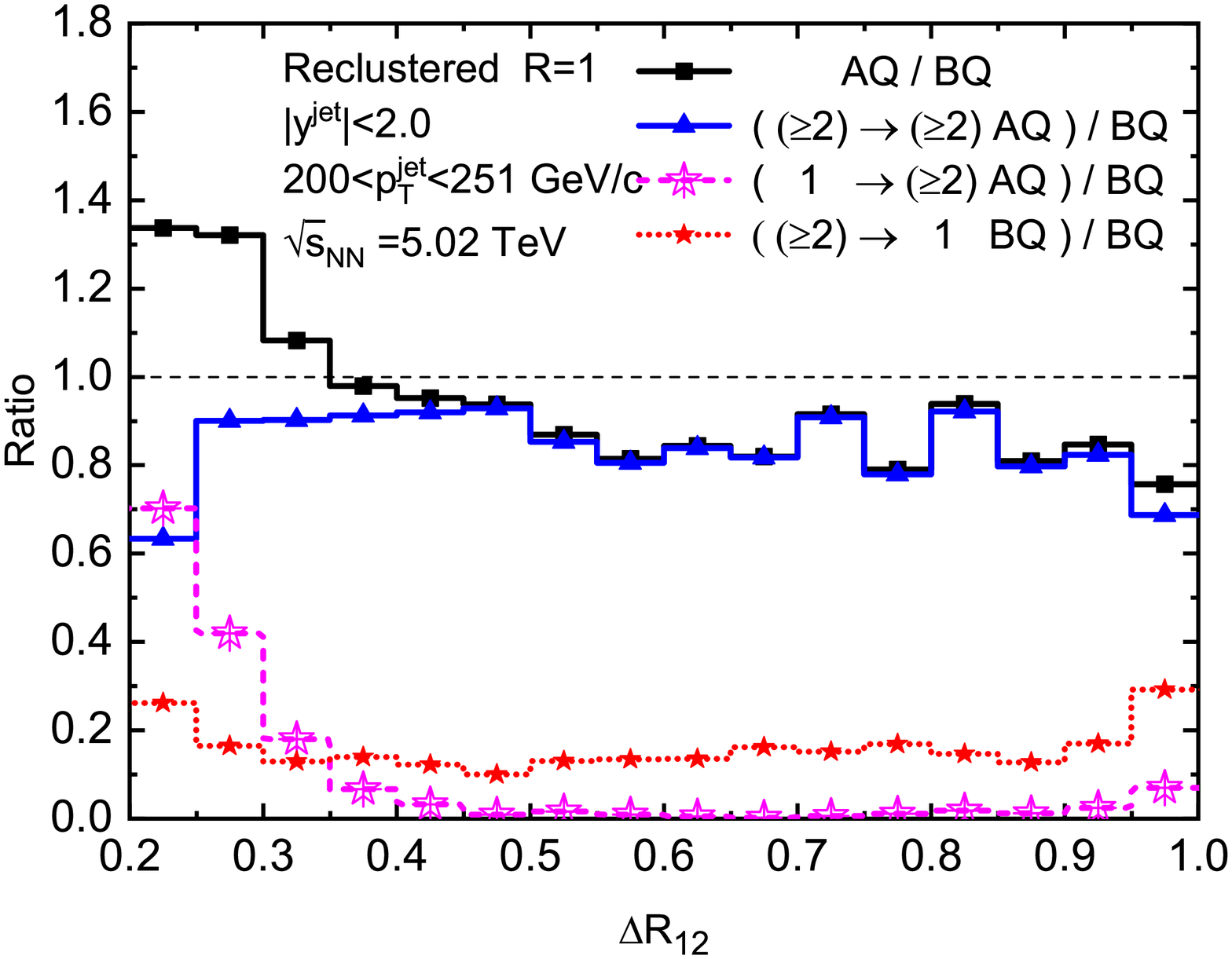}\vspace{-8pt}
  \caption{(Color online)  Contributions  from different processes to reclustered LR jet distributions as a function of the hardest jet splitting angle $\Delta R_{12}$  before jet quenching(BQ) and after jet quenching(AQ) respectively at $\sqrt{s_{NN}}=5.02$ TeV(top). Ratio of distributions from different processes AQ to the distribution BQ(bottom). Detailed descriptions are in the text. }\label{r12_pp_PbPb}
\end{figure}

As can be seen, the $\Delta R_{12}$ in the LR
jets with complex substructure are  mostly focused on small distance region.  An enhancement in small distance region and a suppression in large $\Delta R_{12}$ region are found in Pb+Pb collisions compared to p+p collisions.  It indicates that reclustered LR jets with small splitting angle are less suppressed than those with large splitting angle, which leads to larger value for $R_{AA}$ in small $\sqrt{d_{12}}$ region than that in large $\sqrt{d_{12}}$ region.
The modification factor sharply decreases with $\Delta R_{12}$ for small values of the splitting angle, followed by flattening for larger $\Delta R_{12}$ due to the modification of jet fragmentation pattern, as displayed in Fig.~\ref{r12_pp_PbPb},  which  ultimately will lead to the $\sqrt{d_{12}}$ dependence of $R_{AA}$ in Fig.~\ref{raa_d12}.

\begin{figure}
  \centering
\includegraphics[scale=0.34]{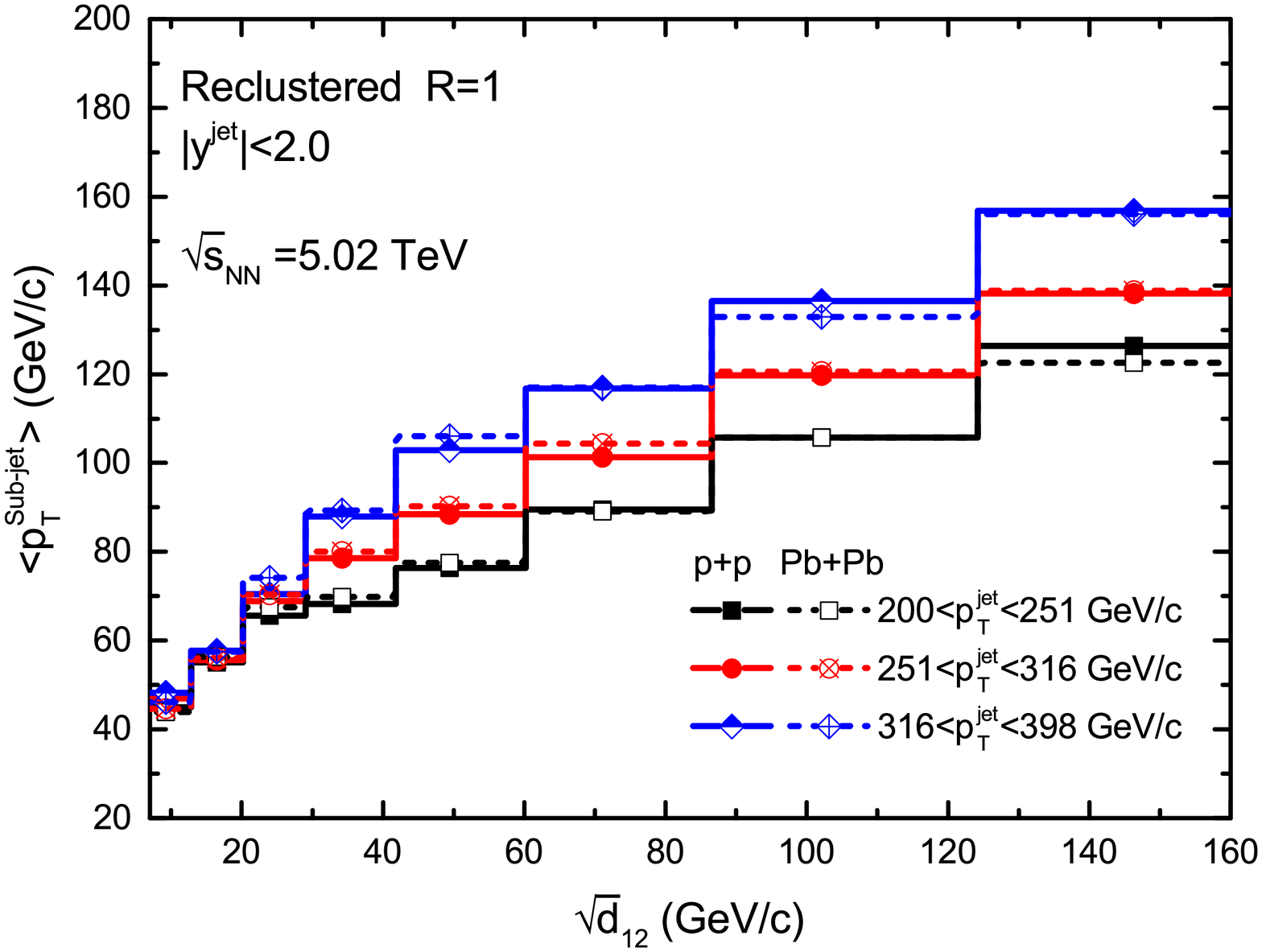}\vspace{-20pt}
\includegraphics[scale=0.34]{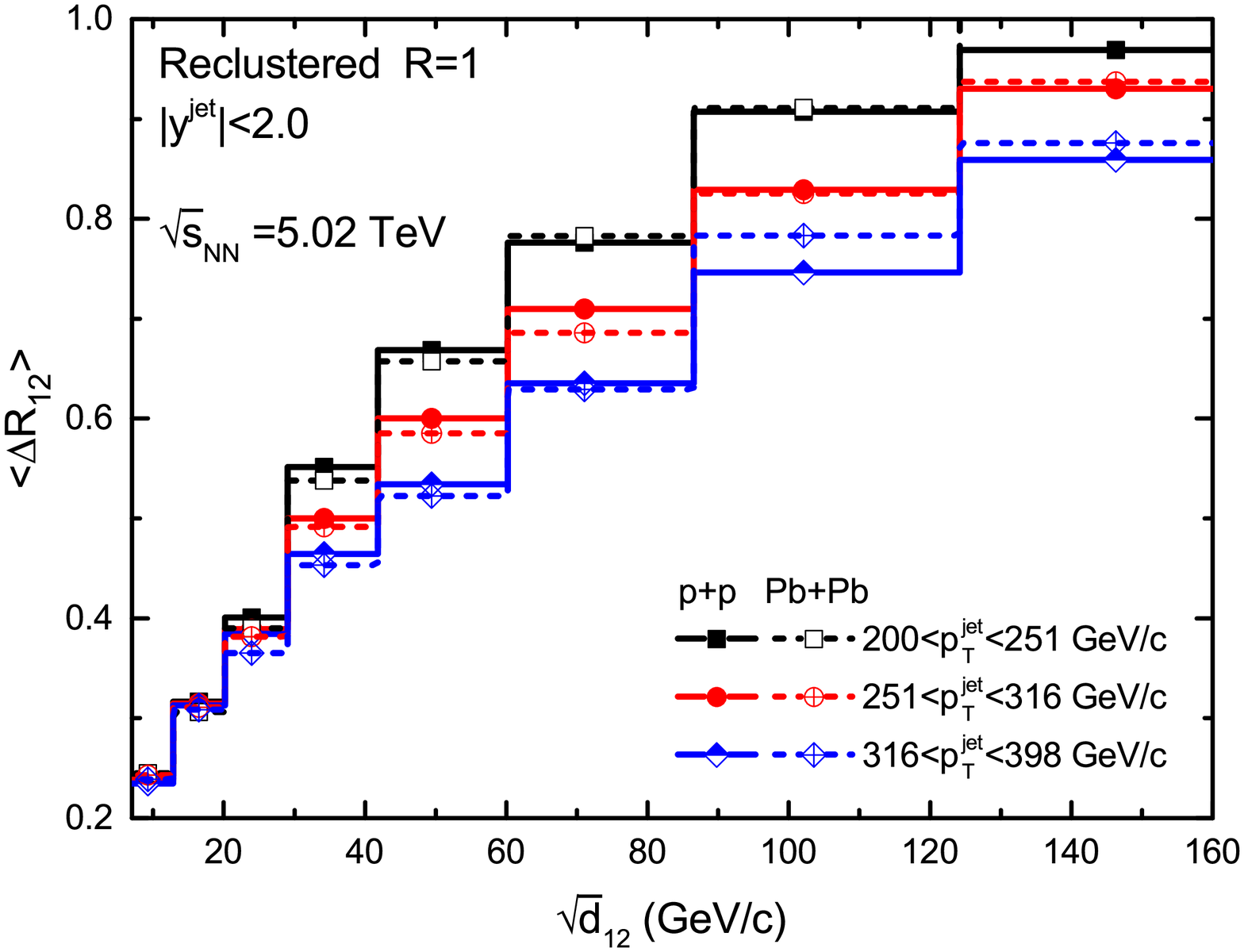}
  \caption{(Color online) Average energy of the hardest splitting  in different reclustered jet transverse momentum intervals in p+p and Pb+Pb collisions as a function of the splitting scale (top panel).  The angle of the hardest splitting  in different reclustered 	LR jet transverse momentum intervals in p+p and Pb+Pb collisions as a function of the splitting scale at $\sqrt{s_{NN}}=5.02$ TeV (bottom panel).   }\label{d12_pt_r12}
\end{figure}

\begin{figure}
\centering
\includegraphics[scale=0.35]{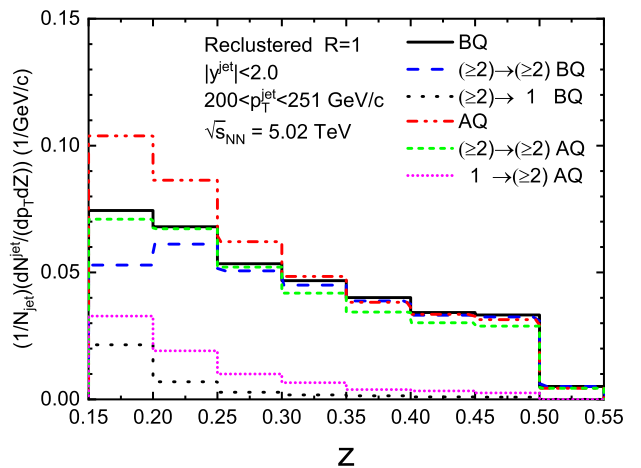}\vspace{-5pt}
  \caption{(Color online)  Contributions  from different processes to reclustered LR jet distributions as a function of the hardest jet splitting function z  of vacuum and medium-modified reclustered LR jets at $\sqrt{s_{NN}}=5.02$ TeV. }\label{z_pp_PbPb}
\end{figure}

In addition,  the modification factor for the normalized fragmentation scale $z$ is  greater than one in small $z$ region and is smaller than unity in large $z$ region as shown in the bottom panel of Fig.~\ref{raa_r12_z}.  The medium modification  of $z$ distribution comes from two sources: the energy shift of  jet and subjet $p_T$ in the QGP, and the medium modifications on the internal structure of  the reclustered LR jet.  We may come back to this in detail in Fig.~\ref{z_pp_PbPb} by presenting more discussions.


 In order to investigate how the hardest splitting is modified due to jet-medium interactions, we need to have a detailed understanding of the structure of the splitting.  The average transverse momentum of soft-subjet $p_T^{Subjet}$(subjet with smaller energy) of the hardest splitting  as well as the angle of the hardest splitting $\Delta R_{12}$ in different reclustered jet transverse momentum intervals in p+p and Pb+Pb collisions as a function of the splitting scale $\sqrt{d_{12}}$ are presented in Fig.~\ref{d12_pt_r12}. We see, both  the average subjet  transverse momentum $\langle p_T^{Subjet}\rangle$ and the distance between the splitting constituents $\langle \Delta R_{12}\rangle$  sharply increase with increasing $\sqrt{ d_{12}}$. There is no surprise that  $\langle p_T^{Subjet}\rangle$ and $ \langle R_{12}\rangle$ are strongly correlated to $\sqrt{ d_{12}}$, however, further detailed analyses show that $\langle p_T^{Sub,jet}\rangle$ is weakly correlated to $\Delta R_{12}$.
 Besides,  almost no difference of the values $\langle p_T^{Subjet}\rangle$ and $\langle \Delta R_{12}\rangle$  between p+p and Pb+Pb collisions with the same jet transverse momentum $p_T^{jet}$ can be observed.   While the $\langle p_T^{Subjet}\rangle$ and $\langle \Delta  R_{12}\rangle$ are quite different in different  $p_T^{jet}$ intervals. $\langle p_T^{Subjet}\rangle$ is much larger for higher energy LR jets, while $\langle \Delta R_{12}\rangle$ is much smaller for larger energy jets.Therefore, it is the change of the yields of the value of $p_T^{Subjet}$ and $\Delta R_{12}$ that lead to the  deviation of the nuclear modification factor from unity in Pb+Pb collisions relative to p+p collisions.



 To further explore the medium modification on the hardest jet splitting pattern,  we calculate  the contributions from different processes to both  vacuum and medium-modified reclustered LR jet distributions as a function of the jet splitting angle $\Delta R_{12}$ and show the results  in Fig.~\ref{r12_pp_PbPb}.  The picture shows the distribution of reclustered LR jets  with $200<p_T^{jet}<251$ GeV/$c$ after propagating through the hot-dense medium. The fragmentation pattern of reclustered LR jets with single subjet(denoted as ``1"), giving zero $\Delta R_{12}$, and multi-subjet( donated as ``$(\ge2)$") processes,giving non-zero $\Delta R_{12}$, are quite different.  The  changes of the fragmentation patterns are separated into four processes: $1 \rightarrow 1$ , $1 \rightarrow (\ge2)$ , $(\ge2) \rightarrow 1 $, $(\ge2) \rightarrow (\ge2)$ due to propagating through the hot-dense medium via LBT.  We don't plot the contributions from $1 \rightarrow 1$ in the following.

 Compared to the total distribution of the LR jets before jet quenching(BQ), the total yield of medium-modified LR jets after jet quenching(AQ) with splitting angle $0.2 <\Delta R_{12}<0.35$  is significantly increased, while is  decreased in the region $\Delta R_{12}>0.35$,  which is similar with the results in Fig.~\ref{raa_r12_z}, indicating that jets with small splitting angle is enhanced, while jets with large splitting angle is suppressed. The modification  patterns are also similar with the results  of JEWEL, QPYTHIA and PYQUEN with wide angle radiation~\cite{Apolinario:2017qay}.

     To illustrate the medium effect on jet structures,
the comparison between jet distribution before (BQ) and after (AQ) jet quenching in $(\ge2) \rightarrow (\ge 2)$ processes is also shown in Fig.~\ref{r12_pp_PbPb}.  The distribution of $(\ge2) \rightarrow (\ge 2)$ AQ  is a little suppressed in region  $0.2 <\Delta R_{12}<0.25$, while is moderately enhanced in the region $0.25 <\Delta R_{12}<0.5$  relative to $(\ge2) \rightarrow (\ge 2)$ BQ. And no distinctness is  observed in large angle region.
 The $\Delta R_{12}$ dependence modification in  $(\ge2) \rightarrow (\ge 2)$ processes is different with the modification pattern of total distribution(or in Fig.~\ref{raa_r12_z}), where large-radius jet yields with large splitting angle jets will be reduced, and jet yields with small splitting angle will be significantly increased in Pb+Pb compared to p+p collisions.

    For further illustration of the modification mechanism, we also calculate other contributions to $\Delta R_{12}$ distributions.
On the one hand,   $1\rightarrow (\ge2$) processes, leading to enahncement of reclustered LR jet yields with $(\ge2)$ after jet quenching,  have considerable contributions to the reclustered  LR jets distributions in small $\Delta R_{12}$ region, especially, more than 50$\%$ in the region $0.2 <\Delta R_{12}<0.25$ . Those processes mainly come from the hard gluon radiation, which will form a new subjet  with other particles due to coherent enhancement. The closer to the jet axis, the more energy will be deposited because of jet-medium interactions. The yields of those processes sharply decrease with increasing $\Delta R_{12}$.

    On the other hand, some subjets will loss energy, leading to that their final transverse momentum fall below the kinematic cut. As a consequence of which,  some reclustered LR jets with complex substructures in vacuum will have only one single subjet  after propagating through the hot-dense medium, like $(\ge2) \rightarrow 1$  process. Those processes will reduce reclustered LR jet yields.  
 The contribution of those processes is widely distributed in the whole space and leads to the reduction of absolute yields of jet with complex substructures due to jet-medium interactions, which therefore gives rise to the suppression of the reclustered LR jets in large splitting  angle $\Delta R_{12}$ region. And the contribution of $1 \rightarrow (\ge 2)$  processes is much larger than that of $(\ge 2) \rightarrow 1$  processes in small angle region, which results in the enhancement of the reclustered LR jets in small splitting angle region.

 \begin{figure}
  \centering
\includegraphics[scale=0.35]{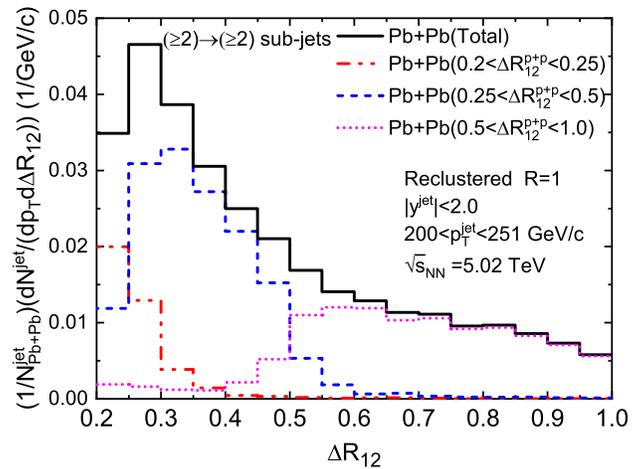}\vspace{-5pt}
  \caption{(Color online)  Medium modification on  the hardest LR jet splitting angle $\Delta R_{12}$   from  $(\ge2) \rightarrow (\ge 2)$ processes to reclustered LR jet distributions as a function of the hardest jet splitting angle $\Delta R_{12}$  of medium-modified reclustered LR  jets at $\sqrt{s_{NN}}=5.02$ TeV. }\label{r12_2_2}
\end{figure}

   To go a step further to see the modification of $(\ge2) \rightarrow (\ge 2)$ processes, we calculate the  contributions from $(\ge2) \rightarrow (\ge 2)$ processes to reclustered LR jet distributions as a function of the hardest jet splitting angle $\Delta R_{12}$  of medium-modified large-radius jets that  originate from different $\Delta R_{12}^{p+p}$ regions and  show the results in Fig.~\ref{r12_2_2}. As mentioned above, large fraction of subjets with very high energy would radiate hard gluons which would form a new subjet due to coherent enhancement.
The splitting angle of those large-radius jets  will be changed due to the reclustering. As can be seen in Fig.~\ref{r12_2_2}, large fraction of jets with splitting angle $0.2<\Delta R_{12}<0.25$ in vacuum will be much more broadened after propagating through the medium. 
    While those jets with splitting angle $\Delta R_{12}>0.5$ will be widely distributed, especially in small $\Delta R_{12}$ region because the radiated subjets are mainly deposited in $0.2<\Delta  R_{12}<0.35$ region. 

Finally, we present the contributions from different processes to reclustered LR jet distributions  as a function of the jet splitting variable $z$ for both  vacuum and medium-modified reclustered LR jets  in Fig.~\ref{z_pp_PbPb}.
The $z$ distribution after jet quenching is moderately enhanced in small $z$ region and  suppressed in large $z$ region in $(\ge2) \rightarrow (\ge 2)$ processes compared to the jets before propagating though the LBT.  To clarify the medium modification of $z$ distribution, we express the splitting variable $z$ in medium as:
 \begin{eqnarray}
   z^{AQ} &=& \frac{p_{T,AQ}^{SubJet}}{p_{T,AQ}^{Jet}}=\frac{p_{T,BQ}^{SubJet}\cdot (1- f^{SubJet}) }
   {p_{T,BQ}^{Jet}\cdot (1- f^{Jet}) }   \nonumber \\
   &=&z^{BQ}\frac{(1- f^{SubJet}) }{(1- f^{Jet}) }
   \label{equation:splitting_z}
 \end{eqnarray}
where $z^{BQ}$ and $z^{AQ}$ are the splitting variable $z$ before and after jet quenching, respectively.  $f^{SubJet}$ and $f^{Jet}$  are energy loss fractions of the reclusted LR jet and its subjet as shown in Fig.~\ref{average_energy_loss}.  In the very small $z$ region with relatively low subjet $p_T$, we have  $f^{SubJet}< f^{Jet}$, thus find $z^{AQ} > z^{BQ}$. Whereas at very large $z$ region with high subjet $p_T$, we observe $f^{SubJet}> f^{Jet}$ (see Fig.~\ref{average_energy_loss}), which then gives $z^{AQ} < z^{BQ}$.

 Furthermore, we study the contribution from $1 \rightarrow (\ge 2)$ processes which increase the jet yields, and $(\ge 2) \rightarrow 1$ processes which reduce the jet yields.   As can be seen, the contribution  from $1 \rightarrow (\ge 2)$ processes is larger than that from $(\ge 2) \rightarrow 1$ processes in the whole region, leading to an overall enhancement of jet yields.

Combining the contributions of medium modifications from three channels: $(\ge 2) \rightarrow (\ge 2)$ processes,  $(\ge 2) \rightarrow 1$ processes, and $1 \rightarrow (\ge 2)$ processes, we find that both the jet energy loss and the change of jet structure will give rise to the enhancement of jet yield in the small $z$ region, leading to large deviation in small $z$ region. While jet energy loss reduces the jet yield in large $z$ region, and  the change of jet structure increases the jet yield. These two mechanisms compete with each other in large $z$ region,  and give a moderate difference in large $z$ region in Pb+Pb collisions compared to p+p collisions,  eventually lead to the $\sqrt{d_{12}}$ dependence of the nuclear  modification factors.

\section{Summary}
\label{summary}

We have carried out the first detailed theoretical investigation of the medium  modification on the reclustered LR jets production as well as  its the hardest parton splitting in Pb+Pb collisions relative to that in p+p collisions. The nuclear modification factor of the reclustered LR jets evaluated as a function of jet transverse momentum is a little smaller than the value of inclusive $R=0.4$ jets. A quantitative calculation of the absolute amount of the transverse momentum missing in the medium shows that reclustered LR jets will lose larger fraction of its energy than inclusive $R=0.4$ jets with the same transverse momentum due to its complex structure. The fraction of energy loss via jet quenching  rapidly increases in the region $p_T^{jet}<80$ GeV/$c$, and smoothly decreases with increasing  $p_T^{jet}$ when $p_T^{jet}>80$ GeV/$c$.   As a result of which,  the nuclear modification factor for reclustered LR jet is a little smaller than that of inclusive $R=0.4$ jet and increases smoothly with increasing $p_T^{jet}$.

The jet spectrum evaluated as a function of the splitting scale $\sqrt{d_{12}}$ of the hardest parton splitting obtained from a reclusting procedure is overall suppressed in Pb+Pb collisions relative to p+p collisions and  the nuclear modification factor $R_{AA}$ sharply decreases with increasing $\sqrt{d_{12}}$ for small values of the splitting scale followed by flattening for larger $\sqrt{d_{12}}$.
The suppression is a result of  the reduction of jet yields as well as the modification on the jet fragmentation pattern.  Jet energy loss dominates the modification in large splitting scale region, while the change of jet  fragmentation pattern via different modification mechanism has almost $50\%$ contributions in small splitting scales region. $\sqrt{d_{12}}$ is strongly correlated to the splitting angle $\Delta R_{12}$ and fragmentation function $z$.   A  detailed calculation of the splitting angle $\Delta R_{12}$ and fragmentation function $z$ shows that reclustered LR jet with small splitting angle or small $z$ is less suppressed, which lead to the $\sqrt{d_{12}}$ dependence  of the nuclear modification pattern.

A further investigation  on splitting angle $\Delta R_{12}$  from $(\ge2) \rightarrow (\ge 2)$ processes shows that jet yield is suppressed in small $\Delta R_{12}<0.25$ region, while is enhanced in $0.25<\Delta R_{12}<0.5$ region, and keeps unmodified in large $\Delta R_{12}$ region.
Further, we find that  $(\ge 2) \rightarrow 1$ processes plays a dominant role in the reduction of jet yields in large $\Delta R_{12}$ region. And the contribution of $1 \rightarrow (\ge 2)$ processes is much larger than that from  $(\ge 2) \rightarrow 1$ processes in small angle region, which gives the enhancement of jet yield in small $\Delta R_{12}$ region.

Finally,  we demonstrate that in $(\ge2) \rightarrow (\ge 2)$ processes the $z$ distribution in Pb+Pb is moderately enhanced in small $z$ region and suppressed in large $z$ region.
In addition, contribution  from $1 \rightarrow (\ge 2)$ processes is larger than that from $(\ge 2) \rightarrow 1$ processes in the whole region, which may moderately increase the jet yields. The total contributions of these three processes give rise to a large deviation in small $z$ region  and  moderate difference in large $z$ region in Pb+Pb collisions as compared to p+p collisions,  and eventually leading to the $\sqrt{d_{12}}$ dependence of the nuclear  modification factors.


{\bf Acknowledgments:}  The authors would like to thank H Zhang, W Dai, T Luo, P Ru,  and G Ma for helpful discussions. This research is supported by Guangdong Major Project of Basic and Applied Basic Research No. 2020B0301030008, Natural Science Foundation of China with Project Nos. 11935007 and 11805167.

\end{document}